\documentclass[a4paper]{article}
\usepackage[normalem]{ulem}
\usepackage{natbib}
\usepackage[usenames]{color}
\definecolor{red}{rgb}{0.9,0,0}
\definecolor{green}{rgb}{0,0.8,0}
\definecolor{blue}{rgb}{0,0,0.8}
\definecolor{cautionred}{rgb}{1.0,0,0}

\newcommand{\half}{\ensuremath{\frac{1}{2}}}
\newcommand{\modu}[1]{\ensuremath{\left| {#1} \right|}}
\newcommand{\magn}[1]{\ensuremath{\modu{#1}^2}}
\newcommand{\peptic}{$\psi$-epistemic }

\title{Quantum- vs. Macro- Realism: What does the Leggett-Garg Inequality actually test?}
\author{O.J.E Maroney\thanks{owen.maroney@philosophy.ox.ac.uk}\\ \textit{Faculty of Philosophy and Wolfson College, Oxford} \and C.G Timpson\thanks{christopher.timpson@bnc.ox.ac.uk}\\ \textit{Faculty of Philosophy and Brasenose College, Oxford}}
\begin{document}
\maketitle

\begin{abstract}\small
\noindent \textit{Macroscopic Realism} is the doctrine that a macroscopic system is always determinately in one or other of the macroscopically distinguishable states available to it, and so is never in a superposition of these states. The Leggett-Garg inequality was derived to allow experimental test of whether or not this doctrine is true. Given their formal analogies, the Leggett-Garg inequality is also often thought of as a temporal version of a Bell-inequality. Notwithstanding a recent surge of interest in the inequality, controversy remains regarding  what would be shown by its violation. Here we resolve this controversy, which in our view arises due to an insufficiently general and model-independent approach to the question so far. We argue that Leggett and Garg's initial characterisation of macroscopic realism does not pick out a particularly natural realist position, but we go on to articulate an operationally well-defined and well-motivated position in its place. We then show in a general setting that much weaker conditions than Leggett and Garg's are sufficient to derive the inequality: in the first instance, its violation only demonstrates that certain measurements fail to be non-disturbing at the operational level. We articulate three distinct species of macroscopic realist position, and argue that it is only the first of these which can be refuted by Leggett-Garg inequality violation. Macroscopic realism \textit{per se} cannot be refuted by violation of the inequality therefore, but this first position is an attractive one, so ruling it out remains of interest. A crucial role is played in Leggett and Garg's argument by the assumption of \textit{noninvasive measurability}. We show that this notion is ambiguous between the weaker notion of disturbance at the operational or statistical level, and the very much stronger notion of invasiveness at the ontic level of properties of the system. Ontic noninvasiveness would be required to rule out macroscopic realism \textit{per se} but this property is not entailed by macroscopic realism, and its presence cannot be established in a model-independent way. It follows that despite the formal parallels,  Bell's and Leggett and Garg's inequalities are not methodologically on a par. We close with some reflections on the implications of our analysis for the pedagogy of quantum superposition.
\end{abstract}

\tableofcontents

\newpage

\section{Introduction}

The Leggett-Garg inequality \citep{lg1985} was introduced as a means of putting to experimental test a world-view which Leggett and Garg called \textit{Macroscopic Realism}. According to this view, and in explicit contrast to what quantum theory allows---indeed, more strongly, in contrast to what quantum theory would sometimes seem to \textit{require}---macroscopic objects must always be in some one determinate macroscopic state or another at any given time. No funny-business of quantum superposition is permitted at the macroscopic level. Leggett-Garg inequalities bear strong formal analogies to Bell-inequalities, except that whereas in a Bell-inequality one considers measurements occuring on two (or more) systems at spacelike separation, in a Leggett-Garg inequality, one considers repeated measurements, at different times, of a single observable, on a single system: a timelike, rather than a spacelike separation between measurements. For this reason, Leggett-Garg inequalities have often come to be called \textit{temporal} Bell-inequalities, and, as with the Bell-inequalities proper, the intention is to rule out by experiment (or at least, to put to experimental test) a broad, interesting, and well-defined class of theories which might seem naturally appealing in some way, but which, on due reflection, have experimental implications which can be shown to conflict with the predictions of quantum mechanics.

Leggett-Garg inequalities have been a source of considerable current interest, having been the subject of a range of new experimental proposals, and, recently, of actual experimental tests \citep{ksgm+2011,DBHJ2011,PMN+2010,WM2010,WJ2008,JKB2006}. Indeed, in the recent experiment of \citet{ksgm+2011}, a test satisfying some of the important original strictures of Leggett and Garg has finally been achieved for the first time.\footnote{It should be noted that \textit{no} test to date would plausibly be construed as involving macroscopic quantities, however. They have all been performed on microscopic systems. However the intention is to provide a proof of principle, with the hope of scaling-up to larger systems in due course. It should also be noted that (perhaps unsurprisingly) in all tests so far, the predictions of quantum mechanics have been borne out. \citet{ELN14} is a comprehensive recent review of various experimental approaches and some theoretical aspects. \citet{guidolg} offers some further analysis of assumptions involved in the Leggett-Garg inequality. } However, notwithstanding the contribution of some careful early commentators \citep{clifton1991,FE1991,EF1992,BGG1994}, there still remains considerable controversy about what exactly would be shown by violation of a Leggett-Garg inequality \citep{Ballentine1987,LG1987,Leggett1988,Leggett2002b,Leggett2002a}. The key question, still in contention, is whether macroscopic realism would in fact be ruled out by experimental violation of the inequality.\footnote{To anticipate: The trouble is not with potential loopholes in the experiment, whether---in parallel to the Bell-case---with experimental imperfections such as noise and detector efficiency (though of course any good experimental implementation and analysis will need to attend to these important features), or with `logical' loopholes such as retrocausality, conspiracy or superdeterminism (`no free will' etc., cf. \citet{bell:freevariables}). Rather, the trouble lies deeper.\label{loopholes}}

We will seek to gain clarity on this question by developing an adequately clear statement of what the position of macroscopic realism actually is. We will also ask how natural---and how interesting---the position thus arrived at might be as a realist response to the quantum behaviour of the micro-world. Our preliminary conclusions will be that the view that Leggett and Garg seem originally to have had in mind---essentially, ordinary quantum mechanics with a macroscopic superselection rule---is not a terribly natural realist position to adopt, and we will compare it with some better-motivated realist alternatives. However, we will also show that by adopting a suitably general operationalist framework as a starting point, one can reformulate the Leggett-Garg inequality in a natural manner, and also go on to introduce a clear and operationally well-motivated macroscopic realist position.

As we shall presently see, in order to derive a Leggett-Garg inequality, one needs not only the assumption of macroscopic realism, but also the assumption of the existence of suitable \textit{noninvasive measurements} \citep{lg1985}. We will clarify the notion of noninvasiveness by distinguishing between two distinct ideas: what we shall call \textit{operational non-disturbance} and \textit{ontic noninvasiveness}, respectively. Ontic noninvasiveness implies operational non-disturbance, but not conversely; whilst operational non-disturbance on its own is sufficient to derive the Leggett-Garg inequality. As we shall explain, therefore, what violation of the Leggett-Garg inequality primarily demonstrates is failure of one's measurements to be suitably operationally non-disturbing.\footnote{A condition related to, but generally weaker than, our notion of operational non-disturbance has previously been discussed in the literature, under the label of `statistical noninvasive measurability' by \citet{clifton1991}, for example, and under the label of `no-signalling in time' by \citet{kb}. As discussed by \citet{guidolg}, a further notion of `marginal selectivity' is deployed in this context by \citet{dk2014}. We will expand on the relations between these notions and our parent notion of operational non-disturbance on another occasion.}

With this analysis in place, and by proceeding within an appropriately general framework, we will go on to demonstrate conclusively that violation of the Leggett-Garg inequality does not imply the falsity of macroscopic realism. We will show that there exist three broad classes of macroscopic realist theories, whilst it is only the first of these three---what we term \textit{operational eigenstate mixture macrorealism}---which cannot allow violation of a Leggett-Garg inequality. The other two classes of macroscopic realist theories each permit violation of the inequality without going against the view that macroscopic quantities must always be determinately of one value or another (equivalently, that macroscopic systems must always be determinately in one macroscopic state or the other). At best, then, experimental violation of a Leggett-Garg inequality does not show the falsity of macroscopic realism \textit{per se}, but only the falsity of operational eigenstate mixture macrorealism. This more narrow result remains far from trivial, however, for, as we shall explain, operational eigenstate mixture macrorealism is quite a natural and an interesting view, and ruling it out indicates that even \textit{macroscopic objects} must behave differently depending on whether or not they are actually being observed.

A recurring theme of our discussion will be the question of the extent to which the case of the Leggett-Garg inequalities really parallels the more familiar case of the Bell-inequalities \textit{methodologically}, as opposed merely to \textit{formally} or \textit{mathematically}. What is crucial to the significance of Bell inequalities is that their violation really does rule out a large, well-defined, and interesting class of theories \textit{by experiment}. In order for this to be the case, the class of theories under test has to be defined in a suitably general manner---one which does not make too many ancilliary theoretical assumptions---and in particular, it needs to be the case that whether or not a given theory does, or ought to, satisfy the crucial conditions, can be determined in a suitably model-independent manner. Thus, in the Bell case, the general framework is one of \textit{beable} theories \citep{bell:localbeables}, against an independently given background of relativistic causal structure, and the crucial condition of \textit{local causality} \citep{bell:localbeables, bell:nouvelle} can both be defined and \textit{motivated} wholly independently of the details of any particular candidate theory which might be under test; in particular it can be defined independently of quantum theory.\footnote{By way of example, this marks an important contrast with another recent question, that of the possibility or otherwise of experimental tests of contextuality, so-called. The standard notion of contextuality derives from the Kochen-Specker theorem \citep{KochenSpecker} but it is \textit{defined} in wholly quantum-mechanical terms. One must assume that a great deal of quantum theory obtains in order even to deploy the definition. Whether a suitably theory-neutral notion of contextuality can be arrived at to give adequate sense to the notion of a true experimental test of whether or not nature is contextual remains a highly controversial question (see \citet{barrettkent2004,spekkens2005a,hermens2011}; and references therein, for discussion).} It is for these reasons that experimental violation of a Bell-inequality shows something significant directly about the world, namely, that no locally-causal theory of the world could be empirically adequate.\footnote{Modulo, of course, some of the niceties to do with loopholes mentioned in fn.\ref{loopholes}. One should also note, of course, that failure of local-causality by no means automatically indicates the \textit{presence} of non-local causation or action-at-a-distance (cf. \citet{erpart1} and \citet{harveybell2014}, for example).}

In our view,  Leggett-Garg inequality violation rules out macroscopic realism \textit{per se} no more than Bell's theorem rules out hidden variables \textit{per se}.  If there is a methodological parallel with Bell's theorem, then, it must be drawn between noninvasive measurability and local causality. However it is our contention that, to date, discussion of the Leggett-Garg inequality has not been pursued in a suitably general and model-independent manner. In particular, discussion surrounding the crucial condition of noninvasive measurability and the question of its relation, if any, to macroscopic realism \textit{per se} has not been, and---we shall argue---\textit{cannot be}, made suitably model independent, so the methodological parallel with Bell's theorem fails.

We close our discussion with some reflections on the relation between tests of the Leggett-Garg inequality and familiar two-slit experiments, and consequent implications for  pedagogy when introducing the concept of quantum superposition.% In an appendix we expand on the relations between our discussion and some previous relevant work on the Leggett-Garg inequality.

\section{The Leggett-Garg Inequality}\label{lg}

Leggett and Garg begin their discussion by noting that:
\begin{quotation}
\noindent``Despite sixty years of schooling in quantum mechanics, most physicists have a very non-quantum-mechanical notion of reality at the macroscopic level, which implicitly makes two assumptions.
\begin{itemize}
\item (A1) Macroscopic realism: A macroscopic system with two or more macroscopically distinct states available to it will at all times \textit{be} in one or other of those states.
\item (A2) Noninvasive measurability at the macroscopic level: It is possible, in principle, to determine the state of the system with arbitrarily small perturbation on its subsequent dynamics.
\end{itemize}
A direct extrapolation of quantum mechanics to the macroscopic level denies this." \citep{lg1985}
\end{quotation}

\noindent They invite us to consider the measurement on a given system---call it $S$---of a quantity $Q$ which can take on one of two distinct values, +1 or -1. If $Q$ is some suitably macroscopic quantity, then it will follow from macroscopic realism (A1) that at any time, the system will have a definite value of $Q$, whether of +1 or -1. Now consider what might happen if one were to make \textit{pairs} of measurements of $Q$ on $S$, with some time-lapse between the measurements. We might, for example, having prepared $S$ in some standard way, measure $Q$ at time $t_{1}$ and then at a later time $t_{2}$. We might instead, following that preparation, have measured $Q$ on $S$ at time $t_{1}$, and then at time $t_{3}$, later than $t_{2}$. Or equally, we might, as another alternative following the initial preparation, have measured $Q$ on $S$ at $t_{2}$ and then again at $t_{3}$. (We may imagine that there is potentially some non-trivial dynamics operating on $S$ as time evolves $t_{1} \leq t \leq t_{3}$.) Call the value of $Q$  at time $t_{i}$, $Q_{i}$.

Now, if the quantity $Q$ \textit{always} takes on a definite value, as will be the case if $Q$ is suitably macroscopic and macroscopic realism obtains, then irrespective of whether or not $Q$ is measured on $S$ at a given time $t_{i}$, $S$ will have a definite value of $Q$ at $t_{i}$, that is, the $Q_{i}$ belonging to $S$ will all be well-defined and of value $\pm 1$ for all $i\in \{1,2,3\}$, whether a measurement is performed at $t_{i}$ or not. Under the assumption that the $Q_{i}$ for $S$ are well-defined for all $i\in \{1,2,3\}$ we can consider certain functions of them, such as the following:
\begin{equation}
Q_{LG}=Q_{1}Q_{2} + Q_{1}Q_{3} + Q_{2}Q_{3}.\label{QLG}
\end{equation}
This quantity $Q_{LG}$ will have different values depending on what values of $Q$ $S$ actually has at the respective times, i.e., it will take on various values for different possible histories of $S$. A little thought reveals that the possible values for $Q_{LG}$ are +3, and $-1$, as follows (see Table~\ref{QLG values}).

\begin{table}[h]
\begin{center}
\begin{tabular}{c|c|c||c}
$Q_1$ & $Q_2$ & $Q_3$ & $Q_{LG}$ \\ \hline
+1 & +1 & +1 & 3 \\
+1 & +1 & -1 & -1 \\
+1 & -1 & +1 & -1 \\
+1 & -1 & -1 & -1 \\
-1 & +1 & +1 & -1 \\
-1 & +1 & -1 & -1 \\
-1 & -1 & +1 & -1 \\
-1 & -1 & -1 & 3
\end{tabular}
\end{center}
\caption{Possible values for $Q_{LG}$.}
\label{QLG values}
\end{table}

Now, how might these possessed values of $Q$ relate to values for $Q$ one might measure in the lab, in particular, to the values that might be obtained in our three distinct pairs of measurement scenarios mentioned above? It is at this point that the assumption of noninvasive measurability, (A2), will come to the fore. Before that, however, let us first assume that our measurements of $Q$ are good measurements, in the sense that if a system $S$ has a definite value of $Q$ at the time of measurement, then this value is accurately revealed in the measurement.\footnote{\citet{clifton1991}, following \citet{redhead:book} calls this property \textit{faithful measurement}.} We will consider making measurements of $Q$ on many copies of our given system, all of which have been subject to exactly the same preparation procedure. Each copy will also be subject to the same time evolutions in the intervals \textit{between} possible measurements: $t < t_{1}, t_{1} < t < t_{2}, t_{2} <t < t_{3}$.  Notice that, in general, even if we have prepared all our systems in the same way, and each has been subject to the same time-evolution before measurement, we may still get a statistical spread of results $\pm1$ when we measure $Q$. This is because, even on the assumption of macroscopic realism, and even given that our measurements are good measurements, it could well be that there are uncontrolled underlying variables at play that pertain to our systems, but which have not been fixed by our preparation procedure. (We shall see much more of this notion as we proceed.) Therefore, among the experimentally accessible and important quantities will be expectation values (evaluated for our ensemble of identically prepared systems) of the form $\langle Q_{i}\rangle$ and $\langle Q_{i}Q_{j}\rangle$. We will use subscripts $M_{i}$ to denote expectation values determined when a measurement of $Q$ at time $t_{i}$ is actually performed. Thus $\langle Q_{1}Q_{2}\rangle_{M_{1}M_{2}}$ denotes the expectation value for the product quantity $Q_{1}Q_{2}$ which is found when a measurement $M_{1}$ of $Q_{1}$ and a subsequent measurement $M_{2}$ of $Q_{2}$ is actually made on each element of the ensemble, and so on. If our measurements are good measurements (and on the assumption that $Q$ values are always definite), then an experimentally determined expectation value $\langle Q_{i}\rangle_{M_{i}}$ is telling us about the statistical spread of \textit{possessed} values of $Q$ in our ensemble at $t_{i}$, and an experimentally determined product expectation value $\langle Q_{i}Q_{j}\rangle_{M_{i}M{j}}$ is telling us about the correlations between the \textit{possessed} values at different times $t_{i}$ and $t_{j}$ in our ensemble.

Suppose we were to perform \textit{three} rather than two successive measurements of $Q$, at times $t_{1}, t_{2}$, and $t_{3}$, on our ensemble. This would allow us to determine experimentally the three quantities $\langle Q_{1}Q_{2}\rangle_{M_{1}M_{2}M_{3}}$, $\langle Q_{1}Q_{3}\rangle_{M_{1}M_{2}M_{3}}$, and $\langle Q_{2}Q_{3}\rangle_{M_{1}M_{2}M_{3}}$. This would allow us, furthermore, to determine the quantity:
\begin{equation}
\langle Q_{LG}\rangle_{M_{1}M_{2}M_{3}} = \langle Q_{1}Q_{2}\rangle_{M_{1}M_{2}M_{3}} + \langle Q_{1}Q_{3}\rangle_{M_{1}M_{2}M_{3}} + \langle Q_{2}Q_{3}\rangle_{M_{1}M_{2}M_{3}}.
\end{equation}
Since the possible values of $Q_{LG}$ for a single system, are, as we know, either +3 or -1, then the expectation value for the ensemble must be bounded by these values:
\begin{equation}
-1 \leq \langle Q_{LG}\rangle_{M_{1}M_{2}M_{3}} \leq 3.\label{bound}
\end{equation}
An important subtlety enters at this point. When all three measurements $M_{1}, M_{2}, M_{3}$ are actually successively performed on each element of the ensemble, then we do not need to assume that $Q$ has definite possessed values prior to each measurement, which are then revealed in measurement, in order to know that inequality~(\ref{bound}) obtains. We can equally well think (mathematically) of the entries in Table~\ref{QLG values} as representing the \textit{outcomes of actually performed measurements}, as opposed to \textit{prior possessed values}. On either understanding of the quantities involved, $Q_{LG}$ will be well-defined, though the \textit{meaning} of the quantities is crucially different in the two cases.

The Leggett-Garg inequality finally comes into view when we introduce the assumption of noninvasive measurability. When all three measurements $M_{1}, M_{2}$ and $M_{3}$ are performed, since each measurement has a definite outcome of either +1 or -1, inequality~(\ref{bound}) \textit{must} always hold. The bound will be satisfied in quantum theory, in any macroscopic realist theory, and indeed in any other theory \textit{at all} which defines joint probabilities of outcomes for a sequence of actually performed measurements! (This is a \textit{very} large class of theories.) But now imagine that both macroscopic realism obtains \textit{and} that it is possible to perform the first and second measurements of $Q$ noninvasively. If both $M_{1}$ and $M_{2}$ are merely, if performed, accurately revealing a previously possessed value, and in so doing, are not at all affecting the future evolution of that value, then it will not make any difference to the possessed values later accurately revealed whether or not these measurements $M_{1}$ or $M_{2}$ were in fact performed. It follows that under these assumptions, the expectation value $\langle Q_{1}Q_{3}\rangle$ experimentally obtained when only $M_{1}$ and $M_{3}$ are performed should be the same as that which would have been obtained if all of $M_{1}, M_{2}$, and $M_{3}$ had been  performed; and similarly, the expectation value $\langle Q_{2}Q_{3}\rangle$ obtained when only $M_{2}$ and $M_{3}$ are performed should be the same as that which  would have been obtained if all of $M_{1}, M_{2}$ and $M_{3}$ had been performed. Finally, by appealing to the fact that $M_{3}$, if performed, is after $M_{2}$, and so it shouldn't make a difference to things that happened before it, one can conclude that there should be no difference in the value of  $\langle Q_{1}Q_{2}\rangle$ experimentally obtained if one were just to perform $M_{1}$ and $M_{2}$, and leave out $M_{3}$.

With these equalities in place, we derive from (\ref{bound}) the conclusion that:
\begin{equation}
-1 \leq \langle Q_{1}Q_{2}\rangle_{M_{1}M_{2}} + \langle Q_{1}Q_{3}\rangle_{M_{1}M_{3}} + \langle Q_{2}Q_{3}\rangle_{M_{2}M_{3}} \leq 3. \label{LGI}
\end{equation}
This is the Leggett-Garg inequality, in one of its standard forms. Unlike the trivial (\ref{bound}) which brings out an entirely straightforward feature of a \textit{single} experimental set-up, (\ref{LGI}) relates in a highly non-trivial way three \textit{distinct} experimental set-ups: the case in which we measure $Q$ at $t_{1}$ and $t_{2}$, the case in which we measure $Q$ at $t_{1}$ and $t_{3}$, and the case in which we measure $Q$ at $t_{2}$ and $t_{3}$. There is no reason at all to suppose that this inequality will hold in general theories, indeed, every reason to suppose it will not. But as we have seen, it \textit{will} be satisfied if the special conditions of macroscopic realism and noninvasive measurability hold.\footnote{Note for completeness that the derivation of (\ref{LGI}) does not fundamentally require that $Q$ be a macroscopic quantity. This only enters if the condition that $Q$ must always have a definite value is to be derived from the assumption of macroscopic realism. One might proceed instead simply by assuming directly that $Q$ always takes on definite values.}  \citet{lg1985} go on to show that it can readily be violated in quantum mechanics, and values less than -1 obtained.

The kind of  physical system which Leggett and Garg originally proposed as an interesting candidate for investigating whether or not their inequality held was an rf SQUID (superconducting quantum interference device), i.e., a ring of superconductor (some millimeters in diameter) with a single Josephson junction, immersed in an external electromagnetic field. In such a device one can have the current in the superconductor circulating clockwise (label it +1) and one can have it circulating anti-clockwise (label it -1). Since the current (of the order of a few microamperes) involves a very large number of charge carriers in motion, it can plausibly be construed as in some relevant sense macroscopic. Quantum mechanics assigns orthogonal states $|+1\rangle$ and $|-1\rangle$ to these two distinct current states, and it will of course allow superpositions of such states: $\alpha |+1\rangle + \beta |-1\rangle$. Macroscopic realism, Leggett and Garg suggest, would require that the SQUID is only ever in one or other of the two distinct current states. The recent tests of the Leggett-Garg inequality mentioned above have explored other kinds of systems.

Leggett and Garg's derivation of their inequality, as above, uses, as we have seen, both the assumption of macroscopic realism, and the assumption of noninvasive measurability. Whilst Leggett and Garg note that noninvasive measurability does not logically follow as a consequence of macroscopic realism, they try to argue that it is nonetheless `extremely natural and plausible' \citep{lg1985} and moreover, they repeatedly assert---more strongly---that it is such a natural corollary of macroscopic realism \textit{per se} that `the latter is virtually meaningless in its absence' \citep[p.R449]{Leggett2002a}, cf. also \citet[p.949]{Leggett1988}. They motivate this thought by inviting us to consider \textit{ideal negative result} or \textit{null-result} measurements.

Suppose that there is some way of measuring $Q$ which proceeds as follows: one has a measuring device which one knows will interact with the target system only if the target system is definitely in some one particular state of $Q$---say, only if $Q=+1$---and it will not interact at all with the target system otherwise. If the system does have $Q=+1$, then the measuring device will reliably indicate this (it is a \textit{good} measuring device, in the sense used before). Then suppose that macroscopic realism obtains and that one's target system to be measured enters this measuring device. If the device registers an outcome, we know that we are accurately revealing a pre-existing possessed value of $Q$ (since we are assuming macroscopic realism, and the measurement is a good one) but our measurement may well disturb that value and its subsequent evolution. However, if the device (assumed to be working properly) \textit{does not} register an outcome, we can infer that the value of $Q$ is -1, for if it had been +1 the device would have registered, yet it did not; and +1 and -1 are the only two options. Yet the value of $Q$ cannot have been disturbed in this measurement, since we know \textit{ex hypothesi} that no interaction took place. This is a \textit{null-result} measurement. The fact that \textit{nothing happened} allows us to infer the value of $Q$, whilst the fact that \textit{nothing happened} also guarantees that our learning this value does not affect the target system, thus does not affect the value of $Q$, or its subsequent evolution.

Notice that in this argument, the possibility of measuring $Q$ noninvasively is derived from the \textit{conjunction} of macroscopic realism with the assumption of the existence of suitable null-result measurements for $Q$. We shall return to examine in detail these arguments surrounding the connection between macroscopic realism and noninvasive measurability.

\section{What is macrorealism? First pass}

We have seen how the Leggett-Garg inequality (\ref{LGI}) is supposed to function as a test for macroscopic realism. With noninvasive measurability a natural corollary of macroscopic realism, and the pair of macroscopic realism and noninvasive measurability together entailing the Leggett-Garg inequality, experimental violation of the inequality would defeat the macroscopic realist picture. But what exactly \textit{is} macroscopic realism, and how realistic a realism is it, in fact? These are the questions to which we now turn.

At first blush, the statement of macroscopic realism (A1) above might seem perfectly clear---given a macroscopic system with two or more macroscopically distinct states available to it, it will at all times be in one or other of those states. But further reflection must give us pause. One might well be concerned with what exactly is meant by, or what would count as being, `macroscopic', whether a macroscopic system, or a macroscopically distinct state. This concern, whilst real enough, is not our immediate object, however. Rather, our first concern is the perhaps more subtle question of what is meant by \textit{state} in `macroscopically distinct states'. What conception of the state of a system are we operating with? Usually, the notion of state comes as part of a theory, or as part of a general framework of theories. What background theory, or framework of theories, are we considering here?

Leggett and Garg's discussion strongly suggests that their background framework is simply quantum mechanical Hilbert space states. \textit{Macroscopic states are the quantum states that one would assign to macroscopic, or collective, degrees of freedom}. Thus, in a SQUID, one does not trouble to assign a (massively entangled) multi-particle quantum state to the \textit{enormous} number of individual microscopic charge-carriers, rather one simply assigns a single state to the collective degree of freedom, the direction of the current, e.g., $|+1\rangle$ or $|-1\rangle$. The content of macroscopic realism is then that the only permissible states of the SQUID are the quantum states $|+1\rangle$ and $|-1\rangle$ (and statistical mixtures thereof), quantum superpositions of these two states being disallowed.

This reading, according to which macroscopic realism is simply quantum theory subject to a macroscopic superselection rule which forbids superposition in certain regimes, emerges perhaps most clearly in the following passage of Leggett's:
\begin{quote}
``...it seems clear that most if not all [criteria for macroscopicness] will have the property that the specific properties of a measuring apparatus as such are irrelevant to its allocation to the `macroscopic' side of the divide; what is relevant is that the different final states of the apparatus are \textit{macroscopically distinguishable}. Thus, we would expect that such theories would have the general feature that Nature, while known to tolerate linear superpositions at the atomic level, \textit{cannot tolerate quantum superpositions of macroscopically distinct states}, whether or not these have anything to do with that small class of physical objects designed by human beings to act as measuring apparatus, but rather always selects a definite macroscopic state. Let us call this hypothesis for brevity `macrorealism'." \citep[p.943]{Leggett1988}
\end{quote}
He goes on:
\begin{quote}
``the hypothesis that Nature does not tolerate linear superpositions of macroscopically distinct states (`macrorealism') is in principle subject to experimental test." \citep[p.944]{Leggett1988}
\end{quote}
It seems that we are in the realm of applying quantum states to describe physical systems, it's just that not all the usual states in the Hilbert space are permitted as physical. (It may be that we do not always get a new physical state by adding together two of the old ones; but the old ones---the  items one is \textit{contemplating} adding together---\textit{are} themselves quantum states.)

However, if macroscopic realism is simply quantum theory with a macroscopic superselection rule, then we must note the following essential point:
\begin{quote}
\textit{There is nothing realist about denying the existence of superpositions, macroscopic or otherwise.}
\end{quote}
Denying the physical possibility of quantum superposition (micro \textit{or} macro) is neither necessary nor sufficient for realism, and it is not necessary in order to account for our determinate experience of the independently existing macroscopic world which surrounds us. Let us flesh this thought out.

To clarify terms, we take realism in the philosophy of science to be the familiar view---in brief---that our theories seek to give us a literally true description of what the world is like, both in its directly observable and its non-directly observable features; that the empirical success of our theories gives us good reason to believe (defeasible, but nonetheless \textit{prima facie} good, reason to believe) that our theories are true or approximately true; and that the statements of our theories are true or false in virtue of mind-independent facts about the world. As applied to quantum theory, realism requires that we interpret the quantum formalism (or at least some significant part of it) as directly representing facts about the physical world. Moreover, since quantum mechanics is a fundamental theory, its scope, according to standard realism, should be considered as universal: it is apt, in principle at least, to describe the \textit{whole} physical world, seamlessly, and all in one go. Furthermore, we should adopt an unassuming physicalism: reference to observers should play no role except in so far as we model these entities as physical systems within the theory;\footnote{Leggett effectively makes this same point in the first part of the quotation above \citep[p.943]{Leggett1988}. Cf. also, famously, \citet{Bell:against}.} and more generally, we should grant big things to be made of little things, and insist that the laws, or other robust generalisations (if any), governing the behaviour of big things be consistent with the laws stipulated for little things: in this instance, consistent with the quantum laws governing the behaviour of the microscopic.

Now: One could deny superposition and yet fail to be a realist simply because one's general view was not a realist one. For example, perhaps one maintains that one's theory (quantum theory with a superselection rule) is not descriptive of \textit{anything} apart from what the results of experiments would be, characterised at the level of the directly observable. (This would be an \textit{instrumentalist} view, according to which one's theory is just an algorithm for organising observable data, rather than a set of claims about how the world fundamentally is below the level of immediate observation). Or again, perhaps one might endorse the truly radical view that mind-independence fails, leading to some form of idealism or phenomenalism. Or one might deny some other component of the standard realist picture.

More significant than this failure of sufficiency, however, is the fact that denial of superposition is not \textit{necessary} for realism: one can seek to incorporate superposition, including macroscopic superposition, into one's realist, descriptive, account of how the mind-independent world is---incorporate it, moreover, in such a way as to recover the determinate nature of our experience, and of the macroscopic world. There are, of course, a number of well-developed approaches to interpreting quantum mechanics which adopt just this approach, and which are fully realist in the ways described above. We will focus initially on two key examples.

Consider first, then, the de Broglie--Bohm theory \citep{deBroglie, bohm:1952, bohmhiley, holland:1995}. This is a theory in which the wavefunction for the total system (up to and including the entire universe) always evolves unitarily (thus superposition, and macroscopic superposition, is rife) but it is supplemented with the specification of definite positions at all times for particles. Momenta for these particles are also well-defined at all times (thus, all particles enjoy continuous and deterministic trajectories), and these momenta are determined by taking the gradient of the phase of the many-body wavefunction; in other words, the wavefunction is directly involved in pushing particles around. This is a deterministic hidden variable theory (the hidden variables are the particle positions, or more generally, configurations for physical degrees of freedom) and it  is empirically equivalent to quantum theory, on the assumption (which may be stipulated or derived) that the probability distribution over initial configurations (understood merely as \textit{ignorance} of the initial configuration)  is given by the Born rule. (This is the assumption of the obtaining of \textit{quantum equilibrium}, as per  \citet{valentini:1991I,valentini:1991II}, for example.) In this theory, the results of measurements are uniquely determined given the initial configuration of the total system, the initial wavefunction, and the interaction Hamiltonian between measuring device and system measured. In this theory, one can have a wavefunction which is in a superposition, yet the values for physical quantities for the system take on definite values. For example, one's system of interest might be in a superposition of being here and there (where \textit{here} and \textit{there} are potentially a macroscopically large distance apart)---that is, the wavefunction has non-zero support in both these two widely distinct regions of configuration space---yet the system is definitely in one or other of the two places, as stipulated by the additional specification of the particle's (or particles') position(s).

The de Broglie--Bohm theory, then, is troubling from the point of view of our initial characterisation, following Leggett and Garg, of macroscopic realism. For this is a theory which allows macroscopic superposition, but yet, at the same time, also allows that the associated macroscopic physical variables take on definite values at all times. We conclude two things immediately, and notice a third.

First, one can evidently be a realist without denying superposition: the de Broglie--Bohm theory illustrates how superposition in the quantum state (including at the macroscopic level) need not lead to indefiniteness of physical variables. Second, and consequently, there must be something wrong with Leggett and Garg's conception of macroscopic realism as being the denial of the possibility of superposition (equivalently, being the assertion of quantum theory with a superselection rule). For, by one perfectly good rendering of the notion of `being in a definite macroscopically distinct state', the de Broglie--Bohm theory will allow that one's system is in a definite (macroscopically distinct) state, viz., being either \textit{here} or \textit{there}, while yet it insists that one's system is also in a quantum superposition. Third, we notice that the de Broglie--Bohm theory will evidently violate a Leggett-Garg inequality whenever quantum theory does, for de Broglie--Bohm theory is empirically equivalent to ordinary quantum theory (given quantum equilibrium obtains), yet it also, on the broader notion of macroscopic realism just delineated (definite values for macroscopic physical quantities), counts as being a macroscopic realist theory. So, the relationship between macroscopic realism and the necessity of satisfying a Leggett-Garg inequality is thrown into doubt.

Now let us consider another very well-developed version of quantum mechanical realism: the Everett interpretation \citep{everett, everettat50, wallace:book}. In this theory, as in de Broglie--Bohm, one entertains a wavefunction for the total system (up to and including the whole universe), again evolving purely unitarily. Here, however, one does not introduce further supplementary quantities, but argues instead that the unitarily evolving wavefunction on its own, construed realistically as directly representing a component of reality (a kind of non-separable field on four-dimensional spacetime, or a separable field on a very much higher-dimensional physical space \citep{spacetimestaterealism}) is enough (given the contingent details of typical dynamics) to underpin emergent macroscopic definiteness: an emergent plurality of effectively non-interacting---thus independent---macroscopic worlds of determinate character; and correspondingly, is enough (on its own) to underpin the emergence of concrete observers, each having definite experiences, within these macroscopically definite (up to decoherence) worlds \citep{wallace:book}.

The Everett interpretation shows us that we can have quantum realism without the denial of superposition. It shows us, moreover, that we can have quantum realism without macroscopic realism, even in the \textit{broader} notion of macroscopic realism which consideration of the de Broglie-Bohm theory encouraged. Everett gives us realism about the macroscopic \textit{without} macroscopic realism---in either the narrow Leggett-Garg sense of quantum mechanics with a superselection rule, or in the broader sense of macroscopic physical quantities always taking on a definite value.\footnote{Everett offers us only the \textit{relative states} being definite, or only \textit{relative quantities} taking on a definite value, not the quantities \textit{tout court}.}

These reflections lead us to proffer a further slogan:
\begin{quote}
\textit{Macroscopic realism is not equivalent to realism about the macroscopic.}
\end{quote}
The de Broglie--Bohm theory shows us that the narrow initial Leggett-Garg characterisation of macroscopic realism as quantum theory with a superselection rule falls short of capturing the full notion of realism about the macroscopic (not to say, realism at all levels of description), whilst the Everett interpretation (in its modern formulations, due primarily to Saunders and Wallace) shows that even the \textit{broadened} notion of macroscopic realism (definite values for macroscopic physical quantities) falls short of capturing the notion of realism about the macroscopic.

Regarding the Leggett-Garg inequality: it is obvious that the Everett interpretation allows violation of the inequality, however at the same time, it delivers determinate macroscopic worlds and determinate experience of those worlds.

An alternative approach to quantum-mechanical realism is to assert the reality of the quantum state, as in de Broglie--Bohm theory and Everett, but to deny that the dynamics of the total system is always purely unitary. Theories in this class are realist wavefunction collapse theories, which postulate some additional term (or terms) in the law describing the evolution of microscopic systems, whose intended effect is to kill-off superposition between terms when it might seem that that's required in order to give a definite result of an experiment---or more generally, to leave the macroscopic world in a determinate (non-superposed) state---but which have very little effect on microscopic superpositions. The Ghirardi--Rimini--Weber (GRW) theory \citep{grw} is a well-known example of such an approach, one of the earliest to be constructed.

Concrete examples of realist collapse theories, such as GRW, will fall under the heading of Leggett and Garg's original notion of macroscopic realism, and indeed, it is quite natural to take this class of theories to be just what Leggett and Garg originally had in mind. In these theories, the additional dynamical terms explain why the superselection rule---no macroscopic superposition---obtains. However, at this point, attention focusses back on what exactly is meant by `macroscopic', and by `macroscopically distinct'. Different ways in which one seeks to implement the collapse mechanism in one's collapse theory will lead to different stories about exactly which kinds of superposition are killed-off. Thus in many ways, Leggett and Garg's `macroscopic realism', as stated, fails to pick-out a natural kind of theories. Which basis (or approximate basis) will one achieve collapse to? Equivalently: with respect to which basis is superposition forbidden (at least on suitable timescales)? These questions can't be answered unless one is presented with the detailed proposed theory.

By way of illustration, consider what the GRW theory would say about Leggett and Garg's favoured example of the distinct macroscopic current states in a SQUID. As was shown by \citet{rae:1990}, even though GRW will quickly entail collapse to determinate states for macroscopic pointer variables of a measuring apparatus (spatially distinct states), it will not entail collapse to one or other of the distinct  $| +1\rangle$ or  $| -1\rangle$ current states of the SQUID. This is because the GRW theory works by spontaneous spatial localisation occuring to microsystems, but since the two distinct current states of the SQUID \textit{wholly overlap} spatially, the localisation process makes almost no difference to either of these states, and certainly doesn't distinguish between them. At best, the effect of GRW localisations will occasionally be to split-up a Cooper pair, and thus increase somewhat the normal dissipativeness of the superconductor; but it will not effect collapse. Thus GRW is a macropscopic realist theory which readily allows that the Leggett-Garg inequality could be violated for measurements on SQUIDs. In general, we will be unable to tell whether a given realist collapse theory will or will not require satisfaction of a Leggett-Garg inequality in a given experimental set-up unless we are told the details of the collapse process. It would seem that if a Leggett-Garg experiment is performed and violation of the inequality is shown, it could still be that a macroscopic realist theory in Leggett and Garg's original sense is the true theory of nature, it is just one which does not impose collapse in the particular basis (or approximate basis) which has been deployed in the experiment. We conclude that so far, Leggett and Garg's macroscopic realism has not, unlike Bell's local causality, picked out a particularly natural class of theories which may be tested directly against experiment.

\section{An operational formulation of the question}

So far we have seen i) that denial of the possibility of macroscopic superposition cannot be the correct way, as it stands, of formulating macroscopic realism; ii) that there can be realist understandings of quantum theory which account for the determinate nature of the macroscopic world which surrounds us and, consequently, which account for the determinate nature of our experience, yet which readily admit (indeed, embrace) macroscopic superposition; and iii) that even having broadened the notion of macroscopic realism away simply from `no macroscopic superposition', one can achieve realism about the macroscopic without having to be a macroscopic realist even in this broader sense. Moreover, we have seen that the question of whether or not macroscopic realism really does forbid violation of the Leggett-Garg inequality is unclear in general; and even in the relatively narrow case of the family of realist collapse theories, whether or not satisfaction of the inequality should hold for a given experimental set-up will depend on the details of the collapse mechanism, which (in principle at least) could vary really quite widely. It is time to go back to the drawing board.

Let us introduce a fairly familiar kind of operational formalism (cf. \citet{spekkens2005a}) as a means of framing the general questions  surrounding macroscopic realism and the Leggett-Garg inequality. Thus we will consider dividing-up any given experimental arrangement in the lab into three components, a \textit{preparation} process, $E$, a \textit{transformation} process, $T$, and a measurement process, $M$, having distinct outcomes taking values $Q=q_{i}$. We consider these three different components, from the point of view of the formalism, simply as black-boxes, with no assumption as to their internal workings. These different elements are to be identified operationally, and in the formalism, the experimental arrangement as a whole is characterised by a probability distribution $P_{(E, T, M)}(Q=q_{i})$, being the probabilities that, given the preparation $E$, followed by the transformation $T$, the measurement $M$ will have such-and-such outcomes. These probabilities are assumed to be measurable from the long-run statistics displayed by the experimental apparatus. Occasionally, for brevity, and where it will not lead to confusion, we may suppress certain indices. Note also that it is somewhat arbitrary how one divides a given experimental arrangement up---one might include a transformation as part of a preparation, for example, rather than treating it as a separate process; equally, one might include a transformation as part of the measurement procedure; or again, one might include a measurement's having had a certain outcome as part of a preparation process, perhaps.

Preparations, transformations and measurements will naturally divide-up into equivalence classes. (Indeed, this is important in the very notion of separating out an experimental arrangement into three general kinds of process.)
\begin{enumerate}
\item Two different preparation procedures $E_{1}$ and $E_{2}$  will be operationally equivalent ($E_{1}\simeq E_{2}$) \textit{iff} for any transformation, and for any measurement, the same probability distribution over outcomes obtains whichever preparation procedure was performed: \[ E_{1}\simeq E_{2} \leftrightarrow \forall q_{i}, T, M\; P_{(E_{1}, T, M)}(Q=q_{i}) = P_{(E_{2}, T, M)}(Q=q_{i}).\]
\item Two different transformation procedures $T_{1}$ and $T_{2}$ are operationally equivalent ($T_{1}\simeq T_{2}$) \textit{iff} for any preparation and for any measurement, it makes no difference to the probabilities for measurement outcomes which transformation took place:
\[ T_{1}\simeq T_{2} \leftrightarrow \forall q_{i}, E, M\; P_{(E,T_{1},M)}(Q=q_{i}) = P_{(E,T_{2},M)}(Q=q_{i}).\]
\item Two different measurement procedures $M_{1}$ and $M_{2}$ will be operationally equivalent ($M_{1}\simeq M_{2}$) \textit{iff} their respective sets of outcomes $Q_{1}=q_{i}$ and $Q_{2} = q_{j}$ can be put into one-to-one correspondence, and whatever the preparation beforehand and whatever the transformation beforehand, there would be agreement on the probabilities assigned to outcomes for these measurements:
\[ M_{1} \simeq M_{2} \leftrightarrow \forall q_{i},E, T\; P_{(E,T,M_{1})}(Q_{1}=q_{i}) =  P_{(E,T,M_{2})}(Q_{2}=q_{i}).\]
In this framework, an equivalence class of measurements naturally corresponds to some physical quantity or other.\footnote{Indeed,  formally, though not metaphysically, one would want to \textit{identify} a physical quantity with an equivalence class of measurement procedures, in this kind of framework.}
\end{enumerate}

Now let us go back to consider the following experimental set-up, familiar from our previous discussion. At time $t_{0}$, a fixed preparation process $E$ is performed on system $S$. At time  $t_{1}> t_{0}$, a measurement $M_{1}$ having two possible outcomes $Q_{1}=\pm 1$ is performed on $S$. At time $t_{2}> t_{1}$, a measurement $M_{2}$, having two possible outcomes $Q_{2}=\pm 1$ is performed on $S$. And finally, at time $t_{3} > t_{2}$, a measurement $M_{3}$, having two possible outcomes $Q_{3}=\pm 1$ is performed on $S$. Between $t_{1}$ and $t_{2}$, $S$ is subject to the time evolution $T_{1}$; and between $t_{2}$ and $t_{3}$, $S$ is subject to the time evolution $T_{2}$.

This whole arrangement is described by the joint probability distribution:
\begin{equation}
P_{(E,M_{1},T_{1},M_{2},T_{2},M_{3})}(Q_{1}=q_{i},Q_{2}=q_{j},Q_{3}=q_{k}).\label{joint}
\end{equation}
Notice that at this stage we have made no assumption that the measurements $M_{1}, M_{2}$ and $M_{3}$ all belong to the same equivalence class, i.e., intuitively, that they are all measurements (perhaps in different ways) of one and the same physical quantity.

When a well-defined joint probability distribution exists, then we can of course derive the probability distribution for a smaller number of the same set of variables, simply by summing-out the remaining variables (i.e., by taking the marginal distribution). For example, if we just wanted to know the probabilities for the outcomes of the first measurement, we could derive it from (\ref{joint}) as follows (suppressing the indices for the initial preparation and the intermediate transformations, for brevity):
\[P_{(M_{1},M_{2},M_{3})}(Q_{1}=q_{i})= \sum_{q_{j}q_{k}}P_{(M_{1},M_{2},M_{3})}(Q_{1}=q_{i},Q_{2}=q_{j},Q_{3}=q_{k}).\]
Similarly, if we were interested in the correlations between the outcomes of the first and second measurement in this set-up, we could calculate them from (\ref{joint}) as follows:
\[P_{(M_{1},M_{2},M_{3})}(Q_{1}=q_{i},Q_{2}=q_{j})=\sum_{q_{k}}P_{(M_{1},M_{2},M_{3})}(Q_{1}=q_{i},Q_{2}=q_{j},Q_{3}=q_{k});\]
and so on for other pairs of variables of interest. Expectation values for the outcomes of the measurements are then simply given by weighting the value of the outcomes by their probability, so, for example:
\begin{equation}
\langle Q_{1}Q_{2}\rangle_{M_{1}M_{2}M_{3}} = \sum_{q_{i}q_{j}} P_{(M_{1},M_{2},M_{3})}(Q_{1}=q_{i},Q_{2}=q_{j})q_{i}q_{j};
\end{equation}
and so on.

Now return to reflect again on Table~\ref{QLG values}, and the quantity $Q_{LG}=Q_{1}Q_{2} + Q_{1}Q_{3} + Q_{2}Q_{3}$. As noted before, when a joint probability distribution for the trio of variables $Q_{1}$, $Q_{2}$, $Q_{3}$ exists, as it does given (\ref{joint}), then evidently
\[-1 \leq \langle Q_{LG}\rangle_{M_{1}M_{2}M_{3}} \leq 3.\]
(Again, notice that here we have not, as we assumed before, required $M_{1}, M_{2}$ and $M_{3}$ all to be measurements of the same quantity, i.e., be in the same equivalence class. We are considering them all to be performed, however.)

As before, we will now seek to compare the behaviour of the experimental set-up where one performs all three of the measurements with the behaviour of the trio of distinct sub-experiments in each of which only two of the stated measurements are performed (though the preparation $E$ and transformations $T_{1}$ and $T_{2}$ remain the same in all three sub-experiments). Of course, in a general operational probabilistic theory in which a joint probability distribution like (\ref{joint}) is defined, we can infer nothing about the joint probabilities for the pairs of values when only two experiments are performed. Since we are considering quite distinct experimental arrangements, the joint probabilities for the pairs of values in experiments in which only two measurements are performed are \textit{not} constrained by the joint distribution for the triples of values when all three measurements are performed. Given that these are distinct experimental arrangements, the former quantities are not to be derived by taking the marginal distributions of the latter.

However, if the following highly non-trivial conditions on the probabilities obtain (again, supressing the indices for the preparation and the transformations), then the Leggett-Garg inequality as in (\ref{LGI}) will follow (though here for the slightly more general case of possibly differing binary measurements $M_{1}$, $M_{2}$ and $M_{3}$):
\begin{eqnarray}
P_{(M_1,M_2)}(Q_{1}\!=\!q_{i},Q_{2}\!=\!q_{j})\!\!\! &=& \!\!\!\sum_{q_{k}} P_{(M_1,M_2,M_3)}(Q_{1}\!=\!q_{i},Q_{2}\!=\!q_{j},Q_{3}\!=\!q_{k}) \label{time order}\\
P_{(M_2,M_3)}(Q_{2}\!=\!q_{j},Q_{3}\!=\!q_{k})\!\!\! &=&\!\!\! \sum_{q_{i}} P_{(M_1,M_2,M_3)}(Q_{1}\!=\!q_{i},Q_{2}\!=\!q_{j},Q_{3}\!=\!q_{k}) \label{M1OPND} \\
P_{(M_1,M_3)}(Q_{1}\!=\!q_{i},Q_{3}\!=\!q_{k})\!\!\! &=&\!\!\! \sum_{q_{j}} P_{(M_1,M_2,M_3)}(Q_{1}\!=\!q_{i},Q_{2}\!=\!q_{j},Q_{3}\!=\!q_{k}). \label{M2OPND}
\end{eqnarray}
It is crucial to notice that, unlike in the standard Leggett-Garg proof, as above, we have here  assumed nothing about the existence or otherwise of definite possessed values of physical quantities prior to measurement. We are working directly with the probabilities alone, which are themselves directly measurable by experiment.

\subsection{Operational non-disturbance implies the Leggett-Garg Inequality}

What do the conditions (\ref{time order}--\ref{M2OPND}) above amount to? They relate the joint probabilities for the case where only \textit{pairs} of measurements are performed to the marginal probabilities for pairs of variables in the case where all \textit{three} measurements are performed.

The first condition, (\ref{time order}), is a very natural, and perhaps obligatory, constraint to do with time order: it says that whether or not one performs a measurement later-on should not affect the probability distribution over the outcomes of previously performed measurements. It is a requirement of \textit{no signalling backwards in time}, so it can be assumed fairly uncontroversially.

The second two conditions, (\ref{M1OPND}) and (\ref{M2OPND}), are stronger than the time-ordering requirement. They are conditions that one's measurements $M_{1}$ and $M_{2}$ are respectively, in a sense to be made more precise in a moment, \textit{operationally non-disturbing}. That is (roughly) it doesn't make any difference to your other statistics whether or not you perform the measurements. This is clearly a very significant assumption to make.

Consider a preparation process $E$, followed by a measurement $M$, followed by a further measurement $M^{\prime}$. As we shall define it, the measurement $M$  is \textit{operationally non-disturbing for the preparation $E$ and subsequent measurement $M^{\prime}$} \textit{iff} it is not possible to tell, based upon the observed statistics of $E$ and $M^{\prime}$ whether or not $M$ was performed. This will hold \textit{iff}:
\begin{equation}
P_{(E,M^{\prime})}(Q^{\prime}=q_{j}) = \sum_{q_{i}}P_{(E,M,M^{\prime})}(Q=q_{i},Q^{\prime}=q_{j}).\label{OPND}
\end{equation}
The case where you do not perform the intermediate measurement is statistically the same as the case in which the intermediate measurement does happen, but you do not observe the outcome.

One can then define stronger notions as follows: $M$ might be operationally non-disturbing, given a prior preparation $E$, for \textit{any} subsequent measurement $M^{\prime}$; $M$ might be operationally non-disturbing for a fixed subsequent measurement $M^{\prime}$ for \textit{any} prior preparation $E$. Most strongly of all, $M$ might be operationally non-disturbing for \textit{any} preparation $E$ and \textit{any} subsequent measurement $M^{\prime}$; in which case it will be said to be operationally non-disturbing \textit{tout court}. This is evidently a very strong constraint indeed.

Notice that it is important in the definition we have given of operational non-disturbance that $E$ might be composed of a sequence of previous processes, including previous measurements, and that $M^{\prime}$ might be composed of a sequence of subsequent processes, including future measurements. Thus there may be a whole sequence of prior and of posterior measurement statistics with respect to which one is assessing whether the presence or absence of $M$ makes a difference. With these possibilities noted, we can now use the notion of operational non-disturbance to clarify the content of our previous conditions (\ref{M1OPND}) and (\ref{M2OPND}).

Condition (\ref{M1OPND}) states that $M_{1}$ is operationally non-disturbing  given the preparation $E$ (index suppressed in the previous sub-section) and the subsequent measurements of $M_2$ followed by $M_3$. Condition (\ref{M2OPND}) states that $M_{2}$ is operationally non-disturbing for the subsequent measurement of $M_{3}$, given the prior preparation process which is constituted by the concatenation of $E$ followed by $M_{1}$'s being performed and having a definite outcome.

Given the unexceptionable time-ordering condition (\ref{time order}), then, a Leggett-Garg inequality immediately follows if the measurements $M_{1}$ and $M_{2}$ are suitably operationally non-disturbing. Violation of the inequality, therefore, shows that one or other, or both, of these measurements must in fact be operationally disturbing. We note that we have had to say nothing here of macroscopic realism, pre-possessed values, or null-result measurements. The logical relation between operational non-disturbance and the Leggett-Garg inequality is simple and direct. Violation of the inequality means that at least one of the first two measurements was operationally disturbing.\footnote{The converse doesn't hold in general. See Appendix for details.}  The crucial question, to which we shall return in due course, is whether or not there is anything about macroscopic realism which suggests that operational non-disturbance should hold for the pertinent measurements.

\section{Operational non-disturbance vs. Ontic non-invasiveness: The Ontic Models framework}

With the condition of operational non-disturbance we have a very simple characterisation of a condition which is sufficent to entail that a Leggett-Garg inequality should be satisfied. Importantly, it is possible to establish directly by experiment whether or not a measurement $M$ is operationally non-disturbing for a given preparation and subsequent measurement, simply by comparing the statistics in the case in which $M$ is performed with those in the case in which it is not.

How does the notion of operational non-disturbance relate to Leggett and Garg's notion of noninvasive measurability which, recall, was crucial to the way that Leggett and Garg derive their inequality (as explained in Section~\ref{lg})? It transpires that two distinct ways of making precise the notion of noninvasive measurability offer themselves. According to the first, noninvasive measurability simply equates to operational non-disturbance. According to the second, it amounts to a very much stronger condition. Both approaches seem to figure in some measure in Leggett and Garg's thought. In order to distinguish the two, we shall need to introduce some further formal apparatus which will provide the context for our discussion in the rest of the paper.

\subsection{The ontic models framework}

The framework we wish to introduce is due to \citet{spekkens2005a} and is known as the \textit{ontological models} or \textit{ontic models} framework. (See also \citet{spekkensharrigan} and \citet{LS2004}.) It bears strong affinities to some ways in which hidden variable theories are often discussed, but it is somewhat more general and is usefully flexible.\footnote{It is somewhat more general, for example, as the framework naturally incorporates versions of quantum theory which are wavefunction realist, but not usually thought of as hidden variable theories.}

The idea is to supplement an operational probabilistic theory of the kind introduced above with some account of what gives rise to the observed probabilities. Thus a system has associated with it an \textit{ontic state}, denoted $\lambda$, belonging to a space of states, $\Lambda$, for the system, where this ontic state captures the real physical properties of the system: properties which the system actually possesses independently of observation or measurement. For example, in $n$-body classical particle mechanics, the ontic state of the system would be the position in phase space; in views of quantum theory which are realist about the quantum state, the quantum state itself would be the ontic state, or part of it.

The framework also associates with a system a probability distribution $\mu(\lambda)$ over the system's set of ontic states $\Lambda$. Thus, in particular, a given preparation process $E$  (in the operational sense above) for a given system will produce a certain probability distribution $\mu_{E}(\lambda)$ over the system's set of possible ontic states ($\int_{\lambda} \mu_{E}(\lambda) d\lambda=1$). If the preparation is very precise, this probability distribution may be a delta function---i.e., one has managed to ensure that the system is definitely in one and only one of its possible ontic states. But it may very well be that no such maximally fine preparations are possible, and that the best one can do is to produce a spread of possible ontic states, with some probability distribution over them. It will be convenient for later-on to introduce the notation \[\textrm{supp}(\mu_{i}) := \{\lambda|  \mu_{i}(\lambda) > 0\}\] for the support of the probability distribution $\mu_{i}(\lambda)$, i.e., for the set of states $\lambda \in \Lambda$ for which the probability distribution $\mu_{i}(\lambda)$ is non-zero. A mixture (convex combination) of preparations $E_{i}$ will give us another preparation: $\sum_{i}w_{i}\mu_{E_{i}}(\lambda)= \mu(\lambda)$, for $w_{i}\geq 1, \sum_{i}w_{i} = 1$.

Here are some familiar examples to illustrate the idea: 1. In classical statistical mechanics with the microcanonical distribution, we have a probability distribution $\mu(\lambda)$ over the ontic states of the system---our system is really lying somewhere on the fixed energy hypersurface, but we don't know where, and each possibility gets equal probability. 2. In realist quantum theory with collapse, our system is always really in some pure state, but it may well be that we don't know which (e.g. following a non-selective measurement process); the system is said to be in a \textit{proper mixture} and we assign a probability distribution over the options.

Let us now consider measurement processes. Measuring devices $M$ \textit{respond} in some way to the system's being in the ontic state $\lambda$ at the time of measurement. This is characterised by the \textit{response function} $\xi_{M}(Q=q|\lambda)$, being the probability that the measuring device $M$ will indicate the outcome value $Q=q$ given that the system is in ontic state $\lambda$ on measurement. In general $Q$ might take on continuous values; for our present purposes we need only consider discrete values $q_{i}$. The response of the measuring device to a particular $\lambda$ might be deterministic, in which case the range of the response function would be $\{0,1\}$, or---more generally---the response might be stochastic, in which case the range of the response function would lie in the full interval $[0,1]$. Evidently, $\sum_{q_{i}} \xi_{M}(Q=q_{i}|\lambda)=1.$  Notice that even if two measurements $M$ and $M^{\prime}$ belong to the same operational equivalence class $(M\simeq M^{\prime})$ they may well not respond in the same way to a given ontic state $\lambda$, i.e., they may have different response functions. In this case, we have \textit{contextuality}, a dependence on the \textit{way} in which a given quantity is measured. (\citet{spekkens2005a} calls this form of contextuality \textit{measurement contextuality}.)

Transformation processes $T$ are represented by stochastic maps from ontic states to ontic states: $\tau_{T}(\lambda|\lambda_{0})$, giving the probability distribution over subsequent ontic states given that one started in the earlier ontic state $\lambda_{0}$. Note that a preparation followed by a transformation gives us another preparation:\[\mu_{(E,T)}(\lambda) = \int d\lambda_{0}\mu_{E}(\lambda_{0})\tau_{T}(\lambda|\lambda_{0}).\]

Putting all of preparation, transformation and measurement together, we now have:
\[P_{(E,T,M)}(Q=q_{i}) = \int d\lambda_{0}d\lambda_{1}\mu_{E}(\lambda_{0})\tau_{T}(\lambda_{1}|\lambda_{0})\xi_{M}(Q=q_{i}|\lambda_{1}).\]

We also wish explicitly to note the effect that performing a measurement, and its having had a particular outcome, may have on the underlying ontic state. We formalise this using a particular transformation map which we associate with the measurement $M$: $\tau_{M}(\lambda|Q=q_{i},\lambda_{0})$. In other words, the probability distribution over ontic states post-measurement may depend on the particular measurement $M$ performed, on the particular outcome $Q=q_{i}$ that occurred, and on what the pre-measurement ontic state $\lambda_{0}$ was. Once more, even if two measurements belong to the same operational equivalence class, it may well be that they do not affect the ontic states of the system in the same way.

As we have said, realist collapse interpretations of quantum theory can be described in the ontic models framework; seeing this will help illustrate how the framework functions. In this case the ontic states $\lambda$ are the pure states $|\psi\rangle \in \mathcal{H}$, where $\mathcal{H}$ is the Hilbert space of the system. If one has a pure-state preparation procedure preparing the system in the state $|\phi\rangle$, then the probability distribution over ontic states generated by the preparation is simply $\mu_{|\phi\rangle}(\lambda)=\delta_{|\phi\rangle,|\psi\rangle}$. A mixed-state preparation is just a convex sum of these. The measurement response functions are as follows, for the simple case of projective measurement. For a measurement $M$ which is associated with a particular orthonormal basis $\{|q_{i}\rangle\}$ of $\mathcal{H}$, we have $\xi_{M}(Q=q_{i}|\lambda_{0}\!=\!|\phi\rangle) = |\langle q_{i}|\phi\rangle|^{2}$. If $M$ is a projective measurement `of the first kind' then $\tau_{M}(\lambda\! =\! |\psi\rangle|Q=q_{i}, \lambda_{0}\!=\!|\phi\rangle) = \delta_{|\psi\rangle,|q_{i}\rangle}$, i.e., the measurement doesn't affect the system if it is already in an eigenstate of the quantity measured, and it leaves it in an eigenstate of the quantity measured otherwise.

A further illustration, which will be useful later-on, is given by the de Broglie--Bohm theory. Here the ontic state of a system is not given by the quantum state $|\psi\rangle$ alone, but by the ordered pair of quantum state and position in configuration space: $\lambda \in \{ (|\psi\rangle, X)\}$.\textit{Now} for a pure-state preparation process, the resultant probability distribution is \[\mu_{|\phi\rangle}(\lambda) = \delta_{|\phi\rangle,|\psi\rangle}|\langle \phi |X|\phi\rangle|^{2}.\] That is, even for the most refined preparation process one can achieve in the theory, there is still a non-trivial (in fact, Born rule) probability distribution over the ontic states. The measurement response functions in the de Broglie--Bohm theory are all \textit{deterministic}, $\xi_{M}(Q=q_{i}|\lambda)\in  \{0,1\}$, but, importantly, they are \textit{contextual}; the value assigned to the outcome of the measurement depends on which particular measurement process it is, not just on the equivalence class that the measurement belongs to.

With the ontic models framework now in hand, let us return to analysing the two distinct notions of noninvasive measurability.

\subsection{Noninvasiveness as operational non-disturbance}

One quite natural way of reading Leggett and Garg's (1985) characterisation of noninvasive measurability at the macroscopic level (A2) is as follows. We are certainly used to thinking of the large-scale macroscopic objects we encounter in our daily lives as being composed of very large numbers of microsystems, the vast majority of whose detailed features make no difference at all to the coarse-grained properties and behaviour of these macroscopic objects which surround us. In a thermodynamic kind of way, the microscopic details wash-out: there are large numbers of different microstates a given object could be in, but exactly which it is in makes no difference at all to the object's large-scale macroscopic features. The space of microstates naturally partitions into distinct sets of microstates, where the same macroscopically observable features obtain whichever of the microstates from a given such set the system happens to be in.

Noninvasive measurability at the macroscopic level might then seem intuitively plausible, since one might suppose that---for all that one may very well disturb the \textit{microstate} of the system when one observes it (exchange of momentum from photons bouncing off, or what have you)---one typically does not change the \textit{macroscopic} features one is measuring, one only moves the microstate around inside the set of macroscopically-equivalent microstates. It might then also seem plausible to assume that the microscopic change one has induced will not affect transition probabilities between macroscopic states of the system in the future. In this case, noninvasive measurability is essentially a statistical notion and is an instance of our notion of operational non-disturbance.

To help formalise this, let us introduce the notion of an \textit{operational eigenstate}. Take an equivalence class of measurements $\tilde{M}=\{M^{\prime}|M^{\prime}\simeq M\}$, and call the corresponding physical quantity, which each $M^{\prime}\in \tilde{M}$ can be thought of as measuring, $\tilde{Q}$. We define an operational eigenstate of $\tilde{Q}$ to be a particular equivalence class of preparation processes. A preparation $E$ is in such an equivalence class \textit{iff}, following $E$, any measurement $M^{\prime}\in \tilde{M}$ would give the same $\{0,1\}$  probability distribution over the outcomes $Q^{\prime}=q_{i}$ of the measurement. That is, if one's system has been prepared in the $q_{i}$ operational eigenstate of $\tilde{Q}$ then any measurement $M^{\prime}\in \tilde{M}$ will return the value $q_{i}$ with probability 1 (and conversely). Now, we know from experience that when the macroscopic properties of a system are actually being observed, the preparation state which the system is thereby put into will be an operational eigenstate of those properties. In other words, having looked at some macroscopic object and seen that it has some particular value for its macroscopic properties, if we look again immediately afterwards (or at any rate, if we look again \textit{soon enough} afterwards), we will find those properties to have the same values.  Operational eigenstates for macroscopic quantities $\tilde{Q}$, therefore, can be used to characterise the behaviour of macroscopic systems while they are being observed.

Measurement of some macroscopic quantity can be performed noninvasively, in the sense articulated above, if there is some measurement $M$ in the operational equivalence class corresponding to the macroscopic quantity in question which is operationally non-disturbing when the system begins in an operational eigenstate of the macroscopic quantity, and given a subsequent measurement of that quantity.

\citet{Leggett1995} seems to have had this statistical notion of noninvasive measurability (an instance of our notion of operational non-disturbance) in mind when he phrased the condition as follows:
\begin{quote}
``(P3) (`noninvasive measurability'): it is in principle possible to measure the value $Q(t)$ on an ensemble without altering the \textit{statistical properties} of that ensemble as regards subsequent measurements." \citep[p.104, emphasis added.]{Leggett1995}
\end{quote}
This formulation would seem to allow that there could be an effect on microscopic features when we are considering a macroscopic system, but that the observable statistics should not change.

Two things to note. First, the requirement of noninvasive measurability, understood in the way we have been suggesting, does not require that \textit{every} way of measuring a quantity  $\tilde{Q}$ be operationally non-disturbing, but only that there be \textit{some} measurement $M$ of $\tilde{Q}$ which is. Second, whilst we have said there is some intuitive plausibility in the story about the possibility of noninvasive measurability and its connection to the notion of the macroscopic (one might make changes at the microscopic level, but these may not affect things at the macroscopic level) this cannot be taken too far. Perhaps in certain circumstances one hasn't noticed anything suggesting that one's observations of macroscopic objects have changed them, or most especially, changed their future behaviour, but very rarely has one actually checked in any detail to see.

\subsection{Noninvasiveness as ontic noninvasiveness}

The second way of understanding noninvasive measurability is as a much stronger condition, what we shall call \textit{ontic noninvasiveness}. A measurement $M$ is ontically noninvasive if it does not change at all the ontic state of the system that obtained prior to measurement, whatever that state might have been. $M$ will be ontically non-invasive for a given outcome $Q=q_{i}$ \textit{iff} the transformation $\tau_{M}$ associated with the process of measurement is as follows: \[\tau_{M}(\lambda|Q=q_{i}, \lambda_{0})=\delta_{\lambda,\lambda_{0}}.\] If $M$ is ontically noninvasive for any outcome value $q_{i}$, then it is ontically noninvasive \textit{tout court}.

Ontic noninvasiveness is a very strong condition on a measurement process. It is evidently enough to entail operational non-disturbance, and thence the Leggett-Garg inequality, as follows. If $M$ is ontically noninvasive then for any preparation $E$ and for any subsequent measurement $M^{\prime}$, there will be no difference to the statistics for $M^{\prime}$ whether or not $M$ was performed, that is, $M$'s being ontically noninvasive entails \textit{complete} operational non-disturbance for $M$ (i.e. operational non-disturbance \textit{tout court}). Therefore $M$ will satisfy the weaker operational non-disturbance conditions (\ref{M1OPND}--\ref{M2OPND}) above (we consider performing the same type of  measurement $M$ at the two different times $t_{1}$ and $t_{2}$), and the Leggett-Garg inequality will follow. But the converse implication doesn't hold. The two weaker operational non-disturbance conditions (\ref{M1OPND}--\ref{M2OPND}) certainly do not entail that $M$ is ontically noninvasive, and even $M$'s being completely operationally non-disturbing does not entail that $M$ is ontically non-invasive, unless it is always possible to prepare arbitrary (and in particular, arbitrarily sharp) probability distributions over one's ontic states, and this will certainly not hold in general in theories. Therefore, a measurement $M$'s being operationally non-disturbing is clearly a significantly logically weaker notion than its being ontically noninvasive.

Now suppose that the measurement $M$ has only two possible outcomes, and that it is only ontically noninvasive for one of these, say $Q=+1$ (the argument extends in the obvious way for the more general case of a larger number of outcomes). If there is another measurement, $M^{\prime}$, in the same operational equivalence class as $M$, which also happens to be ontically noninvasive, but this time only for $Q=-1$, then these two measurement processes, along with a post-selection procedure, can be combined to produce a scenario in which the total effect is that of complete ontic noninvasiveness. The procedure is simple. On an ensemble of identically prepared systems, half the time one should perform $M$ and half the time one should perform $M^{\prime}$. If the outcome of $M$ was +1 one keeps the result, and knows that the ontic state of the system on that run of the experiment was not affected by the measurement; if the outcome of $M^{\prime}$ was -1 one keeps the result, and again knows that the ontic state of the system on that run of the experiment was not affected. In all other cases, the data is discarded. The effect of this post-selection procedure is to select-out a sub-ensemble of the total ensemble where the ontic state of each of the members of the ensemble is guaranteed not to have been affected by the process of measurement. Since $M$ and $M^\prime$ belong to the same operational equivalence class, they can of course be thought of as both measuring the same physical quantity.

Leggett and Garg's central, explicit, argument that noninvasive measurability should hold in the case of macroscopic realism proceeds, recall, by invoking the notion of null-result measurements. The idea is to post-select just those results where no result of the measurement was observed, so where it is believed that nothing happened, so \textit{a fortiori} it is believed that nothing happened to the ontic state of the system on that run, on the assumption that the system must have one or other definite value of the quantity being measured before measurement. In the manner just described, by combining a suitable pair of null-result measurements and post-selecting, the result will be that the ontic states of the systems in one's ensemble, under these assumptions, will not have been affected by the measurement, thus it will be the case that complete ontic noninvasiveness holds for the process as a whole, when one includes post-selection.

The structure of Leggett and Garg's argument from null result measurements shows clearly that they certainly have ontic noninvasiveness in mind as the relevant notion of noninvasiveness, at least at certain crucial junctures of their thought.

\subsection{Summary}

To summarise some of our conclusions from the last two Sections, then:
\begin{enumerate}
\item The specific operational non-disturbance conditions for $M_{1}$ and $M_{2}$, that $M_{1}$ is operationally non-disturbing for the preparation/measurement pair $(E, (M_{2}, M_{3}))$, and that $M_{2}$ is operationally non-disturbing for the preparation/measurement pair $((E,M_{1}), M_{3})$,  suffice (given time-order) to entail the Leggett-Garg inequality:
\[\textrm{OPND}_{\textit{specific}}\longrightarrow \textrm{LGI}.\] N.B. the converse does not hold: satisfying the inequality is not sufficient to show that the measurements are operationally non-disturbing.
\item Complete operational non-disturbance for the measurements $M_{1}$, $M_{2}$, entails specific operational non-disturbance (but not conversely), so:
\[\textrm{OPND}_{\textit{complete}}\longrightarrow \textrm{OPND}_{\textit{specific}}\longrightarrow \textrm{LGI}.\]
\item Ontic noninvasiveness entails complete operational non-disturbance, but not conversely. Measurements which are only ontically noninvasive for one measurement outcome can be combined with post-selection to achieve complete ontic noninvasiveness.
\[\textrm{ONI}\longrightarrow \textrm{OPND}_{\textit{complete}}\longrightarrow \textrm{OPND}_{\textit{specific}}\longrightarrow \textrm{LGI}.\]
\end{enumerate}
It is clear therefore that what Leggett-Garg inequality violation most immediately shows is that one or other, or both, of $M_{1}$ and $M_{2}$ were operationally disturbing for their respective preparations and measurements. These are the weakest conditions that suffice to entail that the inequality should hold. One would need specific reason to believe that one of the stronger conditions in fact held in nature in order for there to be a reason to appeal to (potential) Leggett-Garg inequality violation to rule it out. Notice once more that in none of the chains of inference listed above have we had to invoke macroscopic realism to arrive at the inequality. If we take Leggett and Garg's notion of noninvasiveness simply to be (specific) operational non-disturbance (as we have seen there is some reason to), then the inequality follows directly from that premise alone, without requiring that macroscopic realism holds.  Leggett and Garg suggest that macroscopic realism combined with the existence of null result measurements allows us to infer that the very strong condition of ontic noninvasiveness holds. In that case, violation of the inequality would disprove the \textit{conjunction} of macroscopic realism and the existence of suitable null result measurements. We defer assessing this argument until after we have a workable statement of macroscopic realism on the table.

\section{What is macrorealism? Second pass}

We saw in Section 3 that Leggett and Garg's notion of macroscopic realism did not seem to be capturing any particular natural class of theories which might be considered to instantiate forms of macroscopic realism.  Leggett and Garg's suggestion seems to be based upon the idea of a superselection rule, preventing linear superpositions of macroscopically distinct quantum states.  Unfortunately this suggestion is far too tied to the quantum formalism to make an appropriate model independent criterion in itself.  We will now try and refine the notion in a suitably model independent form.

We approach this through Leggett's comment that ``what is relevant is that the different final
states of the apparatus are \textit{macroscopically distinguishable}''\citep[p.943]{Leggett1988}.  If two states are macroscopically distinguishable, there must be a macroscopically observable difference between them.  A macroscopic realist, then, would believe that these macroscopically observable properties always have determinate values, at all times.

Macroscopically observable properties correspond  in the formalism to the outcomes of macroscopic observations: the chair is in such and such a place, the car is some particular colour.  This suggests rephrasing macroscopic realism as:

\begin{quote}
A macroscopically observable property with two or more distinguishable values available to it will at all times determinately possess one or other of those values.
\end{quote}

The state of the world is captured by the ontic state $\lambda$.  If $M$ is the macroscopic observation of a given property, $\tilde{Q}$, macroscopic realism requires that the response function $\xi_{M}(\tilde{Q}=q_i|\lambda) \in \{0,1\}$ for all possible ontic states.

This must hold for all such measurements, and these measurements must all agree on the value of the macroscopic property: so the physical location of the chair does not depend upon how we look at it.  For any two operationally equivalent macroscopic observations,  $M \simeq M^\prime$, any given ontic state must give the same response: $\forall \lambda \;\; \xi_{M}(\tilde{Q}=q_i|\lambda)=\xi_{M^\prime}(\tilde{Q}=q_i|\lambda)$.

However, the conjunction of these two criteria is well known in quantum theory, as it means that $\tilde{Q}=q_i$ is non-contextually value definite.  This might raise a concern that macroscopic realism is already ruled out by the Kochen-Specker theorem\citep{KochenSpecker}, which forbids any non-contextually value definite formulation of quantum theory.  Fortunately for the macroscopic realist, the Kochen-Specker theorem only entails that \textit{some properties} are contextual\footnote{Though it does entail that this must be the case for every possible ontic state.}.  The macroscopic realist need not be committed to the idea that all the properties of a system are determinate, only that the macroscopically observable ones are.  To avoid confusion, we will use the term `macrodefinite' for `non-contextually value definite for all macroscopically observable properties'.

We have not attempted here to characterise what, exactly, are the macroscopically observable properties, though they may be expected to be a very coarse-grained mutually commuting subset of the quantum observables.

Finally, we may now go on to characterise two ontic states as macroscopically distinct if, and only if, there is at least one macroscopically observable property to which the two states have a zero probability of assigning the same value.  With this definition, macroscopic realism expressed about observable properties recovers macroscopic realism expressed about states.

\section{What does Leggett-Garg Inequality violation in fact rule out?}

The arguments surrounding macroscopic realism and the Leggett--Garg inequality can be boiled down into consideration of the following simplified case. The macroscopic realist regarding a macroscopic quantity $\tilde{Q}$ believes that all ontic states $\lambda$ are non-contextually value-definite for $\tilde{Q}$ (are macrodefinite for $\tilde{Q}$). Now consider a simple experimental sequence of: preparation---measurement of $\tilde{Q}$---measurement of $\tilde{Q}$, i.e., an $(E,M_{1},M_{2})$ sequence, where $M_{1}\simeq M_{2} \in \tilde{M}$, $\tilde{M}$ being the operational equivalence class of measurements which corresponds to the quantity $\tilde{Q}$. (N.B. we of course allow time evolution between $E$ and $M_{1}$ and between $M_{1}$ and $M_{2}$.)

Suppose our macroscopic realist has carefully tested the features of an $M_{1}$ measurement and has established conclusively by experiment that for any operational eigenstate preparations of $\tilde{Q}$, $M_{1}$ is operationally non-disturbing, for any subsequent measurement $M_{2} \in \tilde{M}$.  This amounts to saying that, whenever $\tilde{Q}$ is prepared with a definite value, $M_{1}$ will not disturb the statistics for later measurements of $\tilde{Q}$. Evidently, if a measurement is operationally non-disturbing for a given set of preparations, it will also be non-disturbing for convex combinations of those preparations; so $M_{1}$ is operationally non-disturbing for statistical mixtures of operational eigenstates of $\tilde{Q}$ too.  

Now suppose that we perform the experiment $(E,M_{1},M_{2})$ many times, and compare the resultant statistics with those for the sequence $(E, M_{2})$. (The time evolutions, other than at measurements, are kept the same in the two cases.) Since the macroscopic realist already believes that $M_{1}$ is operationally non-disturbing whenever $\tilde{Q}$ is prepared as having either one value or the other, and since the macroscopic realist believes that $\tilde{Q}$ \textit{must} always take on one value or the other, they will be very surprised indeed if our experiments now reveal that $M_{1}$ was \textit{in fact} disturbing---that the statistics for the two sets of experiments differ after all. How could this be, given that we have already checked that the measurement will \textit{not} be disturbing when we've prepared a definite value for $\tilde{Q}$? Must we conclude that it \textit{cannot in fact be the case} that $\tilde{Q}$ always takes on a definite value, i.e., must we conclude that macroscopic realism is false? If there are ontic states which are not macrodefinite for $\tilde{Q}$ then this would explain how it can be that the measurement was in fact disturbing, when we'd already established that it \textit{would not be} if the systems being measured were prepared as having a definite value of $\tilde{Q}$. This is the core argument in the whole discussion of macroscopic realism.\footnote{If $M_{1}$ is $(E, \tilde{M})$-operationally non-disturbing, then a Leggett--Garg inequality should hold for this simplified arrangement.  We clarify this formally in the Appendix below.}

Suggestive as it is, this argument is not valid as an argument against macroscopic realism, as we shall now see. Careful reflection shows that macroscopic realism can come in a range of sub-varieties, and it is only \textit{one} of these sub-varieties which can be refuted in this way, by showing that one's measurement was after all operationally disturbing---for example, by means of showing that the relevant Leggett--Garg inequality is violated. (Violation of operational non-disturbance in this set-up is a necessary condition for violating the associated Leggett--Garg inequality. See Appendix for details.)

\subsection{Macrorealism 1: Operational Eigenstate Mixture Macrorealism}

The macroscopic realist (for a macroscopic property $\tilde{Q}$) believes that all ontic states $\lambda \in \Lambda$ are macrodefinite for $\tilde{Q}$. Although it is possible to prepare a macroscopic system in such a way that the preparation state is \textit{not} macrodefinite, to the macroscopic realist such a preparation does not mean anything more than that there is an epistemic ignorance of which particular value has been realised by an unknown ontic state.

When the macroscopic quantity is observed, the system is immediately afterwards in an operational eigenstate of the macroscopic quantity, with the observed value.  Given the existence of a measurement procedure, $M_1$, that has been verified to be operationally non-disturbing for any such operational eigenstate, it is not unreasonable to suppose in the general case that the preparation state \textit{prior} to such a measurement was simply the statistical mixture of different operational eigenstates, with different values for the macroscopic property.  The effect of the measurement would then just be to lift the epistemic ignorance of the possessed value, and the post-measurement operational eigenstate would result from a conditionalisation of the probability distribution on the observed value.  Absent any reason to suppose otherwise, this might be seen as the most natural interpretation of the measurement process.

This leads to the following view:

\begin{center}\textit{The only possible preparation states of a system $S$ are operational eigenstates of $\tilde{Q}$ and statistical mixtures thereof.}
\end{center}
Call this view \textit{operational eigenstate mixture macrorealism}. It is clearly a very natural way for a macroscopic realist to think, and it suggests that, whatever may happen at the microscopic level, the macroscopic properties of an unobserved system will continue to behave in the same way as they do when the system is being observed.

Putting things a little more formally, a preparation $E_{q_{i}}$ belongs to the $q_{i}$ operational eigenstate equivalence class $\tilde{E}_{q_{i}}$ \textit{iff} $P_{(E_{q_{i}}, M)}(Q=q_{i}) = 1$, for all $M\in \tilde{M}$. Call the probability distribution prepared by $E_{q_{i}}$, $\mu_{E_{q_{i}}}(\lambda)$. Then for all $\lambda$ and for all $M \in \tilde{M}$, $\lambda \in \textrm{supp}(\mu_{\tilde{E}_{q_{i}}}) \leftrightarrow \xi_{M}(Q=q_{i}|\lambda) = 1$. Denote by $\mu_{q_{i}}(\lambda)$ an arbitrary mixture (convex sum) of operational eigenstate preparation distributions $\mu_{Eq_{i}} (\lambda)$. Then according to the operational eigenstate macrorealist, every preparation state $\mu(\lambda)$ is given by a sum $\sum_{q_{i}} w_{q_{i}}\mu_{q_{i}}(\lambda)$. Every ontic state $\lambda$ in the support of $\mu_{q_{i}}$ is non-contextually value-definite, with value $Q=q_{i}$, and every $\mu_{q_{i}}$ is preparable in the lab, as is every convex sum of such distributions. On this view, the full ontic state space $\Lambda$ is exhausted by the union for all $q_{i}$ of the supports of the distributions $\mu_{q_{i}}$ which arise from the operational eigenstate preparations.

Perhaps the simplest example of operational eigenstate macrorealism is quantum theory with a macroscopic superselection rule.  In this case, the ontic states would be identified with the (pure) quantum state itself, it's just that superpositions of macroscopically distinct states are not physically permitted.  The definition of operational eigenstate mixture macrorealism, therefore, naturally incorporates the type of theory that Leggett and Garg seemed originally to have had in mind.

\sloppypar Now, if operational eigenstate mixture macrorealism obtains, then if one has checked that a measurement $M_{1}$ is operationally non-disturbing for operational eigenstate preparations, it will be $\textit{quite impossible}$ for the statistics for $(E,M_{1},M_{2})$ to differ from those for $(E,M_{2})$, i.e., for $M_{1}$ to be operationally disturbing in this configuration, and consequently for a Leggett-Garg inequality to be violated, for the simple reason that \textit{there are no preparation states available which are not operational eigenstates or mixtures thereof}. Thus should $M_{1}$ prove to be operationally disturbing, either by a direct test, or by violation of the associated Leggett--Garg inequality, we must reject operational eigenstate mixture macrorealism.

\subsection{Macrorealism 2: Operational Eigenstate Support Macrorealism}

Now consider a different view. Suppose it remains the case that every ontic state is macrodefinite for $\tilde{Q}$---so the view is macroscopic realist---and suppose that it remains the case that the full space $\Lambda$ of ontic states is still given by the union of the supports of the distributions which can be arrived at by taking convex sums of operational eigenstate preparations, just as in operational eigenstate mixture macrorealism. Thus no ontic states exist which cannot be accessed by an operational eigenstate preparation (i.e., which do not fall into the support of some $\mu_{Eq_{i}}(\lambda)$). However, suppose that the set of possible probability distributions $\mu(\lambda)$ is larger than the set of convex combinations of operational eigenstate preparation distributions, so that whilst we have the same ontic state space as in operational eigenstate mixture macrorealism, the allowed distributions over that space differ. Call this view \textit{operational eigenstate support macrorealism} (since all the ontic states are in the support of some operational eigenstate preparation.)

Formally, if $\mu_{q_{i}}(\lambda)$ is an arbitrary mixture of operational eigenstate preparation distributions $\mu_{E_{q_{i}}} (\lambda)$, as before, and $\mu(\lambda)$ is an arbitrary preparation of the system, the preparation-support macrorealist believes that $\mu(\lambda)=\sum_{q_{i}} w_{q_{i}}\nu_{q_{i}}(\lambda)$, where $\nu_{q_{i}}(\lambda)>0$ only if $\lambda \in \textrm{supp}(\mu_{\tilde{E}_{q_{i}}})$.  Where the operational eigenstate \textit{support} macrorealist differs from the operational eigenstate \textit{mixture} macrorealist is that for the former, $\nu_{q_{i}}(\lambda)$ cannot in general be expressed as a convex sum of the operational eigenstate preparations $\mu_{E_{q_{i}}} (\lambda)$.

In such a theory it is quite clear how a Leggett-Garg inequality can be violated/a measurement turn out surprisingly to be operationally disturbing, while yet the theory remains fully macrorealist. One might have demonstrated that a measurement $M_{1}$ is operationally non-disturbing for every \textit{operational eigenstate} preparation that has been performed, without it being the case that it is operationally non-disturbing for every preparation $\mu(\lambda)$ that might arise. So for example, even if one starts with an operational eigenstate preparation (or a convex combination of them)---for which one has indeed established that $M_{1}$ would be operationally non-disturbing---it could be that before $M_{1}$ takes place, $\mu(\lambda)$ has evolved away from this initial preparation, and it could well be that the \textit{new} distribution \textit{is} disturbed by $M_{1}$. This would be something that we simply hadn't checked for.  The effect of $M_{1}$ could still be \textit{ontically invasive}: it  may shuffle the ontic states around, so that $\tau_{M}(\lambda|Q=q_{i}, \lambda_{0}) \neq \delta_{\lambda,\lambda_{0}}$, but in such a way that the overall probability distribution does not change, whenever that distribution is an operational eigenstate distribution:
\[
\mu_{E_{q_{i}}}(\lambda)=\int d\lambda_0 \mu_{E_{q_{i}}}(\lambda_0)\tau_{M}(\lambda|Q=q_{i}, \lambda_{0}).
\]  
Operational eigenstates would function rather like equilibrium distributions. However, once the distribution has evolved away from being a mixture of operational eigenstate distributions, then the shifting around of the ontic states which $M_{1}$ induces is such as to disturb the distribution in a way which may be observed:
\[
\nu_{q_{i}}(\lambda) \neq \int d\lambda_0 \nu_{q_{i}}(\lambda_0)\tau_{M}(\lambda|Q=q_{i}, \lambda_{0}).
\]

In operational eigenstate support macrorealism, what allows a Leggett-Garg inequality to be violated is not the existence of ontic states which are not macrodefinite for $\tilde{Q}$, but the existence of novel probability distributions over the ontic states which are not given by mixtures of operational eigenstates. One's measurements of $\tilde{Q}$ might be operationally disturbing for the former, without being operationally disturbing for the latter.

An example of a theory of this kind is given by the original Kochen-Specker non-contextual hidden variable model for a two-state system \citep{KochenSpecker}.\footnote{Granted: it is well known that such a model cannot be extended to higher dimensions, when a sufficient number of distinct, non-compatible, observables are introduced. But this does not entail that such models cannot work for a small enough set of observables. (Macroscopic realism is not the doctrine that one's ontic states must be non-contextually value-definite for \textit{all} quantities.) Indeed, for macroscopic quantities, it is quite plausible that all observable quantities should be \textit{compatible}, so the issue is not, so far, especially troubling.}

\subsection{Macrorealism 3: Supra Eigenstate Support Macrorealism}

In the final variety of macroscopic realism, it of course once more remains the case that all ontic states are macrodefinite for $\tilde{Q}$, but that here, in contrast to both the operational eigenstate mixture and the operational eigenstate support versions of macroscopic realism, there are novel macrodefinite ontic states $\lambda$ which do not fall into the support of any operational eigenstate preparation. When one reliably prepares a system in a definite state of $\tilde{Q}$ (when, that is, one prepares the system in an operational eigenstate) one is only accessing \textit{part} of the macrodefinite-for-$\tilde{Q}$ ontic state space. There are other states $\lambda$ which are still macrodefinite for $\tilde{Q}$ but which may only arise in preparations which have non-zero probabilities for obtaining at least two possible outcomes $Q=q_{i}$ of a measurement $M \in \tilde{M}$: precisely when, in quantum mechanics, superpositions would be prepared.

This view we can call \textit{supra eigenstate support} macrorealism, as there are macrodefinite ontic states which are not contained within the support of any operational eigenstate preparation.  Again, formally, although the supra eigenstate support macrorealist believes that all preparations are of the form $\mu(\lambda)=\sum_{q_{i}} w_{q_{i}}\nu_{q_{i}}(\lambda)$, where $\nu_{q_{i}}(\lambda)>0 \rightarrow \xi_{M}(Q=q_{i}|\lambda) = 1$, the supra eigenstate support macrorealist also believes there exist $\lambda \notin \textrm{supp}(\mu_{\tilde{E}_{q_{i}}})$ for which $\nu_{q_{i}}(\lambda)>0$.

We saw earlier that the de Broglie--Bohm theory caused considerable trouble for Leggett and Garg's discussion of macroscopic realism and violation of the Leggett-Garg inequality, since the de Broglie-Bohm theory certainly seemed, by one perfectly good measure, to be macroscopic realist, but yet it allowed violation of the Leggett-Garg inequality just as ordinary quantum mechanics does. The de Broglie--Bohm theory is precisely an example of supra eigenstate support macrorealism, where the macrosopic quantity $\tilde{Q}$ is (possibly coarse-grained) position. Every ontic state $\lambda=(|\psi\rangle,X)$ is non-contextually value-definite for position, but when $|\psi\rangle = \alpha|q\rangle + \beta|q^{\prime}\rangle$, the distribution $\mu_\psi(\lambda)$ over ontic states again cannot be given as a convex sum of the operational eigenstate distributions, $\mu_q(\lambda)$ and $\mu_{q^{\prime}}(\lambda)$, because novel---but still macrodefinite---ontic states are being accessed. The Leggett-Garg inequality can readily be violated in the de Broglie--Bohm theory, then, as, once more, when checking that our measurement $M_{1}$ was operationally non-disturbing for operational eigenstate preparations, we simply have not checked that our measurement $M_{1}$ is operationally non-disturbing for all distributions over macrodefinite ontic states that there can be.\footnote{\citet{guidolg} also discusses the case of the de Brogile--Bohm theory in the context of Leggett-Garg violation.}

\section{Macrorealism and null result measurements: Macrorealism does not imply non-invasiveness}

Let us at long last come to consider in detail Leggett and Garg's arguments that macroscopic realism entails---or at least very strongly supports---noninvasiveness, in the strong sense of ontic noninvasiveness. If this were so, then Leggett-Garg inequality violation really would cause trouble for macroscopic realism \textit{per se}, rather than merely for the specific form of operational eigenstate mixture macrorealism. A significant part of our argument so far has been that weaker conditions than macroscopic realism \textit{per se} are sufficient on their own to entail the Leggett-Garg inequality so that macroscopic realism itself need not be impugned by violation of the inequality. Again, if Leggett and Garg's argument for ontic noninvasiveness works, then since ontic noninvasiveness entails the weaker conditions, which entail the Leggett-Garg inequality, macroscopic realism itself would be in danger, rather than (merely) the logically weaker conditions.

The argument for ontic noninvasiveness as a corollary of macroscopic realism goes as follows. Suppose we have available a pair of null-result measurements for $\tilde{Q}$: we believe that $M_{1}^{\prime}\in \tilde{M}$ will interact with our system $S$ \textit{iff} $S$ has a definite value $+1$ (say) of $\tilde{Q}$, and we believe that $M_{2}^{\prime\prime} \in \tilde{M}$ will interact with $S$ \textit{iff} $S$ has a definite value $-1$ of $\tilde{Q}$. If macroscopic realism obtains (but not otherwise), then when $M_{1}^{\prime}$ does not fire, we may infer that the value of $\tilde{Q}$ was $-1$, and that there can have been no affect on $S$, since there was no interaction. \textit{Mutatis mutandis} for when $M_{1}^{\prime\prime}$ does not fire. Then by post-selection on cases of no-firing, we can be assured of complete ontic noninvasiveness.

Let us tighten this argument up. In the framework we have adopted it is not possible to formalise the notion of a null-result measurement directly, in part for the very good reason that it is not at all a notion on which one may get direct operational grip. That one might believe a measurement $M_{1}^\prime$ only to involve interaction with $S$ in certain circumstances depends on what one's detailed model of the physical interaction is. Different physical models of the interaction might disagree about whether that was the case.

Nevertheless, one might postulate as follows. Suppose we simply stipulate that $M_{1}^\prime \in \tilde{M}$ is such that every $\lambda$ which is non-contextually value-definite for the -1 outcome is not affected by $M_{1}^\prime$, and that $M_{1}^{\prime\prime}$ is such that every $\lambda$ which is non-contextually value-definite for the +1 outcome is not affected by $M_{1}^{\prime\prime}$. If (and only if) every ontic state is macrodefinite for $\pm 1$, then following post-selection on the -1 outcome for $M_{1}^{\prime}$ and on the +1 outcome for $M_{1}^{\prime\prime}$, we will have a process which overall is guaranteed to be completely ontically noninvasive. We do not talk of a `null-result' here, as any measurement $M\in\tilde{M}$ is always symbolised as having a result $Q=\pm 1$; however we can still call these measurements the representations in our formalism of what Leggett and Garg have in mind. So we have shown that in our formalism, with this characterisation of null result measurements, macroscopic realism and the existence of null result measurements like $M_{1}^{\prime}$ and $M_{2}^{\prime\prime}$ do indeed entail complete ontic noninvasiveness, and thence on down the chain:
\[ (\textrm{MR and NRM})\rightarrow \textrm{ONI} \rightarrow \textrm{OPND}_{\textit{complete}}\rightarrow \textrm{OPND}_{\textit{specific}}\rightarrow \textrm{LGI}.\]

Of course, one immediate remark to make on this argument is that ontic noninvasiveness does not follow from macroscopic realism alone, but only from macroscopic realism with the additional \textit{stipulation} of measurements which are ontically noninvasive for given outcomes for certain of the macrodefinite ontic states. And that there should exist such measurements is clearly no part at all of the notion of macroscopic realism. The claim that all ontic states are macrodefinite does not require, in order for it to be sure of meaning, that some kinds of measurements of the macroscopic quantity should be partially ontically noninvasive (\textit{pace} \citet[p. 949]{Leggett1988}, \citet[p.R449]{Leggett2002b}).\footnote{This claim would only follow according to a curiously extreme positivist conception of empirical meaning.} To be sure, and for all its importance, we do not feel it likely that Leggett or Garg would be much inclined to dispute this point. Rather, it seems to us, their main thought is that \textit{given} that one has a set of measurements in mind which would function as null-result measurements, then it would be all but impossible to maintain macroscopic realism if those measurements were to turn out to be operationally disturbing. This is of a piece with the idea that one cannot hope to turn to just \textit{any old} set of measurements, even of macroscopic quantities, in order to set-up an interesting test of the Leggett-Garg inequalities: one had better have some good reason to believe in the first place that the macroscopic realist will think that the measurements ought to be operationally non-disturbing, for to repeat, their claim is $\textit{not}$ that \textit{every} measurement need be noninvasive, whether in the sense of operationally non-disturbing or in the sense of ontically noninvasive. If one can find \textit{some} measurements that the macroscopic realist might be inclined to think should be noninvasive, then one's Leggett-Garg test can begin to get off the ground.

However, what remains as a point of fundamental importance is that there is nothing \textit{model independent} that can be appealed-to to establish whether or not one should think of one's measurements as being partially ontically non-invasive in the way described above. If one happens to have certain views as to how the detailed physics of the interaction between system and measuring apparatus goes, then one might very well believe that one had a pair of measurements apt for null-result, ontically non-invasive, measurement. But it could well be that one's model is wrong, rather than that it is macroscopic realism which is at fault. Here we find a very fundamental difference with the case of the Bell inequalities. That one's theory should be locally causal followed in a model-independent way from the setting of relativistic causal structure: spacelike separation automatically motivated a certain kind of independence condition. Nothing of the sort holds here---there is no general principle, or independent court of appeal, which might suggest that one's measurements should be partially ontically non-invasive. This can only be a model-dependent hypothesis.

\citet{LG1987} and \citet[p.950]{Leggett1988} suggest that it may be possible to get an independent grip on the required property of ontic noninvasiveness directly by experiment `at least up to a point' \citep[p.950]{Leggett1988}: they consider (what is in our terminology) testing operational  non-disturbance for operational eigenstate preparations. It is certainly true that this is an essential move when beginning discussion of when a macroscopic realist should believe a Leggett-Garg inequality should hold. Without first settling that the measurements in question are operationally non-disturbing for operational eigenstates \textit{no} macroscopic realist is compelled to believe that a Leggett-Garg inequality should follow. But as we have seen, establishing such operational non-disturbance falls enormously short of establishing anything like ontic noninvasiveness, and whilst the operational eigenstate mixture macrorealist would be defeated if a subsequent test were then to show operational disturbance, we have seen that there are two other respectable macroscopic realist positions which have no trouble at all in incorporating operational disturbance for measurements which have previously been shown to be operationally non-disturbing for operational eigenstates.

In a telling passage, Leggett and Garg maintain:
\begin{quote}
``Should anyone wish to interpret the results of our proposed experiment (assumed for present purposes to agree with QM) by saying that the macroscopic object is indeed in a definite macroscopic state but is \textit{nevertheless physically affected by an interaction which we know could have occurred only if it had been in a} different \textit{macroscopic state} he or she is free to do so; we leave it to the reader to judge whether such an interpretation in any way diminishes the force of the quantum measurement paradox, or the significance of our proposed experiment." \citep[emphasis added.]{LG1987}
\end{quote}
But the crucial point is: how would one know what one is supposed to know here, if the point is to carry any force? How do we \textit{know} that an interaction could have occurred only if the system had been in the other state? (How do we know that the measurement was ontically noninvasive for the outcome recorded?) The answer is that we cannot, in a model independent manner. We can only \textit{assume}, and our assumption may well be wrong.

\section{Conclusions}

We have seen that macroscopic realism should be understood not as the claim that certain kinds of quantum superposition are not possible, but as the claim that all ontic states are non-contextually value-definite for a macroscopically observable quantity $\tilde{Q}$. We have shown that macroscopic realism would not be impugned by a Leggett-Garg inequality violation involving measurements of $\tilde{Q}$. Within the notion of macroscopic realism \textit{per se} we have seen that there are three distinct broad kinds of theories: operational eigenstate mixture macrorealism, operational eigenstate support macrorealism, and supra eigenstate support macrorealism. It is only the first of these which is unable to account for potential Leggett-Garg inequality violation.

The remaining two positions do not come without some price however.  A measurement that has been verified to be operationally non-disturbing for operational eigenstate preparations, turns out to be disturbing for preparations which are not operational eigenstates.  The operational eigenstate support macroscopic realist must suppose that the measurement was perturbing the ontic states underlying the operational eigenstate preparations after all, but in such a way as previously to have been operationally undetectable.  The supra eigenstate support macrorealist may continue to believe that the ontic states in the support of eigenstate preparations were not disturbed by the measurement, but at the cost of having to introduce novel, albeit macrodefinite, ontic states in those circumstances when (according to quantum theory) superpositions are prepared.  Neither option is particularly welcome, at least if the appeal of macroscopic realism was to be that the macroscopic properties of things continue to behave in the same way when they are unobserved as they do when they are observed. Leggett-Garg inequality violation therefore remains an interesting result, even if it fails to rule out macroscopic realism \textit{per se}.

Regarding the Leggett-Garg inequality itself we showed, first, that the inequality could be derived directly from some simple conditions pertaining to whether or not one's measurements were operationally disturbing.  This derivation needed to make no mention of macroscopic realism, nor of any notion of noninvasiveness stronger than non-disturbance at the statistical level.  We analysed Leggett and Garg's arguments from the possibility of null-result measurements to their conclusion, that noninvasiveness in a stronger sense than operational non-disturbance was a natural corollary of the notion of macroscopic realism, and found that no such argument could be maintained without appeal to potentially tendentious and model-dependent assumptions.  It follows that, for all their mathematical similarity, the Leggett-Garg inequalities and the Bell-inequalities are not methodologically on a par. Unlike local causality, ontic noninvasiveness cannot be motivated as a general feature which should exist for at least some measurements when macroscopic realism obtains.

\newpage
\section{Coda: The Leggett-Garg Inequality and the Two-slit experiment}

When it comes to introducing the concept of quantum superposition in lectures, one almost invariably begins by appealing to the two-slit experiment, or related interference phenomena, to argue that something distinctive and non-classical must be happening. One tends to argue, roughly, that interference effects show that one cannot understand what is happening in a two-slit experiment as the quantum particle determinately going through one slit or determinately going through the other, and we are just ignorant of which, since then we ought to have a statistical mixture of the two cases, which is not what one sees on the detecting screen. This then motivates the idea that in these experiments, the particle must be enjoying some new way of being---dubbed \textit{superposition}---in which it is not the case that it determinately goes through one, or it determinately goes through the other, slit. Perhaps one says that the particle's spatial path through the slits is indeterminate, or  indefinite, or undefined, or that in some novel sense the particle goes through \textit{both} slits (but not by splitting in two!), or somesuch. But whatever one says by way of positive characterisation of superposition, the main point is that the observed interference pattern shows that the particle is not determinately taking one or other of the options open to it. Slit-position for the particle is not a value-definite quantity in these experiments, one concludes.

This argument is a venerable one, and it is also a very powerful and persuasive one. Nevertheless, it is possible that one might feel a slight twinge of awkwardness or intellectual dishonesty when presenting it. One might hear a small, quiet voice inside saying---how watertight is that argument really? Isn't it just  a little bit \textit{too} quick? How sure are we \textit{really} that one couldn't cook up some kind of theory in which the particle always determinately went through one slit or the other, but managed even so to give the correct interference-effect statistics? One might even have the de Broglie-Bohm theory in mind as an explicit example according to which the particle always really does go through one or the other slit, yet one will still see the correct interference pattern.

Now, as others have also noted before (e.g., \citet{BGG1994}), there are in fact interesting connections between the Leggett-Garg inequality and these venerable interference experiments.  Violation of the Leggett-Garg inequality in quantum theory involves preparing superposition states, and the refutation of operational eigenstate mixture macrorealism, which such violation entails, can be read as the implication that such superpositions cannot be understood as statistical mixtures of eigenstates.  These are therefore the same problems which are presented in the traditional analysis of the two-slit experiment. But we can now put things the other way around. Working backwards from our detailed analysis of the Leggett-Garg inequality we can ask: What may one truly infer from the two-slit experiment, if one is going to be careful about it? 

\subsection{The two-slit experiment}

In the experiment, a quantum system at time $t=t_0$ is prepared in a state, $\psi(x,t_0)$, a wavepacket moving orthogonally to the $x$ direction, towards a barrier containing two narrow slits.  When the wavepacket reaches the screen, at time $t=t_1$, its width in the $x$ direction is much greater than the distance between the two slits, and it may reasonably be represented as a plane wave.  The (renormalised) portion of the wavepacket passing through the first slit is $\psi_1(x,t_1)$ and through the second slit $\psi_2(x,t_1)$.  Finally the two wavepackets spread out on the other side of the barrier, before reaching a detection screen, at time $t=t_2$, where the $x$-position is observed.  $\psi_1(x,t_1)$ evolves into $\psi_1(x,t_2)$ and $\psi_2(x,t_1)$ evolves into $\psi_2(x,t_2)$

The relative frequency with which the quantum object is detected at some particular value of $x$ is given by $\half \magn{\psi_1(x,t_2)+\psi_2(x,t_2)}$.

If the first slit is blocked, the overall intensity is reduced by a half, and the relative frequency becomes $\magn{\psi_2(x,t_2)}$, while if the second slit is blocked, this becomes $\magn{\psi_1(x,t_2)}$.  The problem is then traditionally presented in these terms: if the quantum object passes through the first slit, it can't know whether or not the second slit is blocked, and so it should still arrive at the screen with probability $\magn{\psi_1(x,t_2)}$.  But equally if it passes through the second slit it would be unaware of the status of the first slit, and so it should arrive at the screen with probability $\magn{\psi_2(x,t_2)}$.  So, given it has equal probability of passing through either slit, when both slits are open, the detection probability should be $\half\left(\magn{\psi_1(x,t_2)}+\magn{\psi_2(x,t_2)}\right)$ and not $\half \magn{\psi_1(x,t_2)+\psi_2(x,t_2)}$.  The two terms differ by the quantity $\modu{\psi_1(x,t_2)}\modu{\psi_2(x,t_2)}\cos\left(\phi(x)\right)$, where $\phi(x)$ is the difference in phase between $\psi_1(x,t_2)$ and $\psi_2(x,t_2)$ at the location $x$.

If we now analyse this in terms of the Leggett--Garg inequality, the situation corresponds to the simplified arrangement given at the end of the Appendix. (To emphasise: quantum states $\psi$ are here being taken to label equivalence classes of preparation procedures, rather than ontic states.)  The initial state, $\psi(x,t_0)$ can be treated as an eigenstate of the operator $Q_1$ with eigenvalue $+1$.  For the intermediate measurement, $Q_2$, $\psi_1(x,t_1)$ is assigned the eigenvalue $+1$, corresponding to the system going through the first slit, and $\psi_2(x,t_1)$ is assigned $-1$, corresponding to the system going through the second slit. That is, $\psi_1(x,t_1)$ and $\psi_2(x,t_1)$ are operational eigenstates for measurement of `which slit?'.  Finally, an arbitrary location $x$ on the screen is given the value $+1$ for the final measurement, $Q_3$.

Under these value assignments,
 \begin{eqnarray*}
P_{(M_1,M_2,M_3)}(Q_{1}\!=\!+1,Q_{2}\!=\!+1,Q_{3}\!=\!+1)&=&\half\magn{\psi_1(x,t_2)} \\ P_{(M_1,M_2,M_3)}(Q_{1}\!=\!+1,Q_{2}\!=\!-1,Q_{3}\!=\!+1)&=&\half\magn{\psi_2(x,t_2)} \\ P_{(M_1,M_3)}(Q_{1}\!=\!+1,Q_{3}\!=\!+1)&=&\half \magn{\psi_1(x,t_2)+\psi_2(x,t_2)}
\end{eqnarray*}
  The disturbance due to the intermediate measurement can then be quantified as the difference between the sum of the first two of these and the last:
  \[
  D_{2,+1,+1}=\modu{\psi_1(x,t_2)}\modu{\psi_2(x,t_2)}\cos\left(\phi(x)\right).
  \] It is straightforward to check that this results in the following Leggett--Garg inequality:
\begin{equation}
\langle Q^+\rangle_{LG} = 2 \modu{\psi_1(x,t_2)}\left(\modu{\psi_1(x,t_2)}+\modu{\psi_2(x,t_2)}\cos\left(\phi(x)\right)\right)-1.
\end{equation}
(See Appendix for the general definition of disturbance measures of this type, and how the corresponding Leggett-Garg inequalities follow.)
From this it follows that violation of the inequality $\langle Q^+\rangle_{LG} \geq -1$ requires $\cos\left(\phi(x)\right)< -\modu{\psi_1(x,t_2)}/\modu{\psi_2(x,t_2)}$.\footnote{An equivalent inequality can be violated for $\modu{\psi_1(x,t_2)}>\modu{\psi_2(x,t_2)}$ by a different value assignment for the intermediate measurement.}

The equivalent condition to macroscopic realism, in this context, is that there is always a matter of fact about which slit the quantum object passes through, even with a superposition preparation.  Operational non-disturbance of the operational eigenstates of $Q_{2}$ (i.e., of slit-position for the system) requires verifying experimentally that, if the system is prepared in the state $\psi_1(x,t_2)$, no measurement localised in the vicinity of the second slit will affect the intensity on the screen, which is indeed the case. The same holds for preparation in the state $\psi_2(x,t_2)$ and measurement at the first slit. (Recall that we have to establish operational non-disturbance for the operational eigenstate preparations first in order to be able to go on to reach any interesting conclusions whatsoever.)  Null result (thus \textit{ontically} noninvasive) measurements correspond to measuring only one slit at a time, then discarding the runs in which the system was detected at that slit, and this effect is achieved exactly by blocking one or other of the slits.

This analysis of the two slit experiment, in the terms of the Leggett-Garg inequality, both formalises and clarifies the normal analysis of quantum interferometry experiments. And as we now know, there is more than one kind of view available which would insist that the system must always go through one slit or the other:\footnote{ We put the options in terms of our three flavours of macroscopic realism, but for analysis of the two-slit experiment, there is no need that the quantities involved be macroscopic, of course. The point is rather that the quantities should be non-contextually value-definite, which, as we have seen, is what a macroscopic realist requires of macroscopic quantities.}

\begin{enumerate}

\item In \textit{operational eigenstate mixture macrorealism}, each operational eigenstate associated with a given slit corresponds to a distribution over a set of non-contextually value-definite ontic states, and the superposition preparation of the interference experiment must be described as a statistical mixture of these distributions.  Such a position cannot account for the interference pattern or the attendant Leggett-Garg inequality violation.  Violation of the operational non-disturbance condition is sufficient to refute operational eigenstate mixture macrorealism, therefore, and this happens whenever the interference term is non-negligible.

However, to repeat, operational eigenstate mixture macrorealism is not the only way to believe that the quantum object always passes through one slit or the other.  For the condition of operational non-disturbance to follow for superposition preparations, it is also necessary to believe that measurements performed at one slit cannot be influencing the future behaviour of quantum systems passing through the remote slit, even under these different preparation conditions.

\item \textit{Operational eigenstate support macrocrealism} holds that the superposition preparation only accesses the same ontic states as the operational eigenstates of slit position do, but that the introduction of the measurement must disturb these ontic states, even in the case of null result measurements.  So, even when the system is prepared to be in $\psi_1(x,t_1)$, the introduction of a detector, or a blockage, in the second slit is in fact disturbing the ontic state of the system, even if, with that preparation, one doesn't see any evidence of that disturbance in terms of a shift in the pattern on the detecting screen.

\item \textit{Supra eigenstate support macrorealism} allows that the ontic states prepared by $\psi_1(x)$ are not disturbed by a detector in front of the second slit.  However, the cost is to introduce novel ontic states for the preparation $(\psi_1(x) + \psi_2(x))/\sqrt{2}$, which are available neither to $\psi_1(x)$ nor to $\psi_2(x)$.  These novel ontic states are all non-contextually value-definite, in that they are associated with passing through one or the other slit, and reveal which on measurement. But these novel ontic states all  share the feature that any such state which has the system passing through the first slit will be disturbed by a detector placed in front of the second slit, and vice versa.

\end{enumerate}

So our venerable argument \textit{was} too quick: there are two general families of theories according to which a system definitely always goes through one slit or the other, yet where an interference effect is displayed---i.e., where it makes a difference what is going on at the other slit. The inference from i) there being preparations which are not equivalent to a statistical mixture of operational eigenstates for slit-position, to  ii) there is not a fact about which slit the system went through, is invalid. However, with that said, one can also see the appeal of the venerable argument more clearly. In order to account for the experimental results, the macroscopic realist---or the believer in `always definitely one slit or the other!'---is forced, in one form or another, to accept that a quantum object that is associated with passing through one slit, must nevertheless be affected by a detector placed in the other\footnote{It is interesting that this is not restricted to detectors.  Equivalent conclusions, by simply introducing a phase shift in front of one of the slits, can be obtained from Hardy's ontic state indifference theorem \citep{Hardy2013}, noting that preparation support macrorealism is a form of \peptic interpretation of quantum theory.  The implications for supra eigenstate support  macrorealism also find parallels in the `surrealistic trajectories' arguments surrounding de Broglie--Bohm theory (\citet{ESSW}, \citet[Chpt. 3]{Maroney}.}.  The macroscopic realist cannot entirely avoid the conclusion that, even if something goes through only one slit,  its future behaviour can have a lawlike dependence on what happend at the other slit.\footnote{With the two slit case, because one is dealing with a spatial separation between the slits, it might be tempting to appeal to locality to motivate the thought that in this instance---if not in general---we have good reason to suppose that a measurement at one slit should be partially ontically noninvasive: should not affect a system located at the far slit. Locality is certainly a strong component in the intuitive pull of the venerable argument. But notice that this cannot be taken too far: there is no requirement that goings-on at one slit immediately affect ontic states which are value-definite for the other slit. It is the \textit{future} behaviour of the system, in the overlap of the future light-cones of the two slits, which is affected, and this is consistent with a local transmission of the influence---whatever it may be---from happenings at the slit where the system wasn't.}

In some discussions of the two-slit experiment, one finds claims along the lines that no \textit{plausible} theory could be cooked-up which recovers the quantum predictions whilst insisting that the particle only ever went through one or other slit. But this forces one to enquire: what makes for plausibility? What is hidden here? Our analysis, based on the Leggett-Garg inequality and our three-way distinction between macroscopic realist positions, notes the strength of the traditional argument, but it also shows exactly what space there is for alternative accounts of interference effects which maintain the idea that the particle always determinately goes through one slit or the other, whilst it also clearly flags the costs of these approaches. In this there is considerable advantage. We have moved from the sphere of the often implicit, the potentially subjective, and the regularly tendentious---i.e., judgements of what is physically plausible---to a clear and well-defined space of theories having certain explicit features.

What, finally, of the pedagogy, though? Well, no doubt there is considerable merit in sticking with the traditional line---at least for initial pedagogical purposes: an initial, first-order approximation. But our analysis of how the Leggett-Garg considerations apply in this case shows where one can  immediately point the discerning student, for higher-order corrections, or for an explanation of what might, in these cases, be hidden behind the little (possibly weasel?) word `plausible'.

\newpage

\begin{appendix}

\section{Appendix: The role of operationally detectable disturbance in violating the Leggett Garg Inequality}\label{disturbance}

To draw out the role of operationally detectable disturbances in violating the Leggett Garg Inequality, it helps to introduce a few simple expressions.  Starting with the marginal statistics for the experimental arrangement when all three measurements $M_1,M_2,M_3$ are performed:
\begin{eqnarray}
P_{(M_1,M_2,M_3)}(Q_{1}\!=\!q_{i},Q_{2}\!=\!q_{j})\!\!\! &=& \!\!\!\sum_{q_{k}} P_{(M_1,M_2,M_3)}(Q_{1}\!=\!q_{i},Q_{2}\!=\!q_{j},Q_{3}\!=\!q_{k}) \label{M3mean}\\
P_{(M_1,M_2,M_3)}(Q_{2}\!=\!q_{j},Q_{3}\!=\!q_{k})\!\!\! &=&\!\!\! \sum_{q_{i}} P_{(M_1,M_2,M_3)}(Q_{1}\!=\!q_{i},Q_{2}\!=\!q_{j},Q_{3}\!=\!q_{k}) \label{M1mean} \\
P_{(M_1,M_2,M_3)}(Q_{1}\!=\!q_{i},Q_{3}\!=\!q_{k})\!\!\! &=&\!\!\! \sum_{q_{j}} P_{(M_1,M_2,M_3)}(Q_{1}\!=\!q_{i},Q_{2}\!=\!q_{j},Q_{3}\!=\!q_{k}). \label{M2mean}
\end{eqnarray}
it is useful to define:
\begin{eqnarray}
D_{1,q_j,q_k}&=&P_{(M_2,M_3)}(Q_{2}\!=\!q_{j},Q_{3}\!=\!q_{k})
-P_{(M_1,M_2,M_3)}(Q_{2}\!=\!q_{j},Q_{3}\!=\!q_{k})\label{D1}\nonumber \\
D_{2,q_i,q_k}&=&P_{(M_1,M_3)}(Q_{1}\!=\!q_{i},Q_{3}\!=\!q_{k})
-P_{(M_1,M_2,M_3)}(Q_{1}\!=\!q_{i},Q_{3}\!=\!q_{k})\label{D2}\nonumber \\
\end{eqnarray}
to quantify the change in marginal statistics of $M_2,M_3$, depending on whether $M_1$ is measured, and in the marginal statistics of $M_1,M_3$ depending one whether $M_2$ is measured.

If the introduction of the $M_1$ measurement cannot be detectable by an observer who only has access to the marginal statistics of $M_2,M_3$, then all of $D_{1,q_j,q_k}=0$.  Similarly, if the introduction of the $M_2$ measurement is undetectable with only the marginal statistics of $M_1,M_3$, then all of $D_{2,q_i,q_k}=0$.  An equivalently defined $D_{3,q_i,q_j}$ must all be zero to avoid signalling backwards in time.

Now, looking at each term in the Leggett Garg Inequality, when all three measurements are performed:
\begin{eqnarray}
\langle Q_{1}Q_{2}\rangle_{M_{1}M_{2}M_{3}} &=& P_{(M_1,M_2,M_3)}(Q_{1}\!=\!+1,Q_{2}\!=\!+1)+P_{(M_1,M_2,M_3)}(Q_{1}\!=\!-1,Q_{2}\!=\!-1) \nonumber \\
 && -P_{(M_1,M_2,M_3)}(Q_{1}\!=\!+1,Q_{2}\!=\!-1)-P_{(M_1,M_2,M_3)}(Q_{1}\!=\!-1,Q_{2}\!=\!+1) \nonumber \\
\langle Q_{1}Q_{3}\rangle_{M_{1}M_{2}M_{3}} &=& P_{(M_1,M_2,M_3)}(Q_{1}\!=\!+1,Q_{3}\!=\!+1)+P_{(M_1,M_2,M_3)}(Q_{1}\!=\!-1,Q_{3}\!=\!-1) \nonumber \\
 && -P_{(M_1,M_2,M_3)}(Q_{1}\!=\!+1,Q_{3}\!=\!-1)-P_{(M_1,M_2,M_3)}(Q_{1}\!=\!-1,Q_{3}\!=\!+1) \nonumber \\
\langle Q_{2}Q_{3}\rangle_{M_{1}M_{2}M_{3}} &=& P_{(M_1,M_2,M_3)}(Q_{2}\!=\!+1,Q_{3}\!=\!+1)+P_{(M_1,M_2,M_3)}(Q_{2}\!=\!-1,Q_{3}\!=\!-1) \nonumber \\
 && -P_{(M_1,M_2,M_3)}(Q_{2}\!=\!+1,Q_{3}\!=\!-1)-P_{(M_1,M_2,M_3)}(Q_{2}\!=\!-1,Q_{3}\!=\!+1) \nonumber \\
\end{eqnarray}
A straightforward rearrangement yields the very simple form
\begin{eqnarray}
\langle Q \rangle_{M_{1}M_{2}M_{3}} &=& 4 \left( P_{(M_1,M_2,M_3)}(Q_{1}\!=\!+1,Q_{2}\!=\!+1,Q_{3}\!=\!+1)+ \right. \nonumber \\
 && \left. P_{(M_1,M_2,M_3)}(Q_{1}\!=\!-1,Q_{2}\!=\!-1,Q_{3}\!=\!-1) \right) -1
\end{eqnarray}
showing clearly that the Leggett-Garg Inequality cannot be violated.

When only pairs of measurements are performed:
\begin{eqnarray}
\langle Q_{1}Q_{2}\rangle_{M_{1}M_{2}} &=& P_{(M_1,M_2)}(Q_{1}\!=\!+1,Q_{2}\!=\!+1)+P_{(M_1,M_2)}(Q_{1}\!=\!-1,Q_{2}\!=\!-1) \nonumber \\
 && -P_{(M_1,M_2)}(Q_{1}\!=\!+1,Q_{2}\!=\!-1)-P_{(M_1,M_2)}(Q_{1}\!=\!-1,Q_{2}\!=\!+1) \nonumber \\
\langle Q_{1}Q_{3}\rangle_{M_{1}M_{3}} &=& P_{(M_1,M_3)}(Q_{1}\!=\!+1,Q_{3}\!=\!+1)+P_{(M_1,M_3)}(Q_{1}\!=\!-1,Q_{3}\!=\!-1) \nonumber \\
 && -P_{(M_1,M_3)}(Q_{1}\!=\!+1,Q_{3}\!=\!-1)-P_{(M_1,M_3)}(Q_{1}\!=\!-1,Q_{3}\!=\!+1) \nonumber \\
\langle Q_{2}Q_{3}\rangle_{M_{2}M_{3}} &=& P_{(M_2,M_3)}(Q_{2}\!=\!+1,Q_{3}\!=\!+1)+P_{(M_2,M_3)}(Q_{2}\!=\!-1,Q_{3}\!=\!-1) \nonumber \\
 && -P_{(M_2,M_3)}(Q_{2}\!=\!+1,Q_{3}\!=\!-1)-P_{(M_2,M_3)}(Q_{2}\!=\!-1,Q_{3}\!=\!+1) \nonumber \\
\end{eqnarray}
and it can now be easily seen that if
\begin{equation}
\langle Q\rangle_{LG} = \langle Q_{1}Q_{2}\rangle_{M_{1}M_{2}} +\langle Q_{1}Q_{3}\rangle_{M_{1}M_{3}} +\langle Q_{2}Q_{3}\rangle_{M_{2}M_{3}}
\end{equation}
then
\begin{equation}
\langle Q\rangle_{LG} -\langle Q\rangle_{M_{1}M_{2}M_{3}}= \left(\sum_{q_j=q_k} D_{1,q_j,q_k}+
\sum_{q_j=q_k} D_{2,q_i,q_k}\right) - \left(\sum_{q_j \neq q_k} D_{1,q_j,q_k}+
\sum_{q_j \neq q_k} D_{2,q_i,q_k}\right)
\end{equation}
This can be simplified by noting that $\sum_{q_j,q_k} D_{1,q_j,q_k}=\sum_{q_i,q_k} D_{2,q_i,q_k}=0$, and
it follows that
\begin{eqnarray}
\langle Q\rangle_{LG} &=& 4 \left( P_{(M_1,M_2,M_3)}(Q_{1}\!=\!+1,Q_{2}\!=\!+1,Q_{3}\!=\!+1)+P_{(M_1,M_2,M_3)}(Q_{1}\!=\!-1,Q_{2}\!=\!-1,Q_{3}\!=\!-1) \right) \nonumber \\
    &   & + 2 \left(\sum_{q_j=q_k} D_{1,q_j,q_k}+\sum_{q_j=q_k} D_{2,q_i,q_k}\right) -1
\end{eqnarray}
For the Leggett Garg Inequality to be violated it is a necessary condition that at least one of the set $D_{1,q_j,q_k},D_{2,q_i,q_k}$ be non-zero\footnote{It is not a sufficient condition. $D_i,q_j,q_k$ may be non-zero (and hence operational eigenstate mixture macrorealism be ruled out, see Section 7.1) even if the associated Leggett--Garg Inequality holds.}.  At least one of the $M_1$ or $M_2$ measurements must introduce a disturbance which is detectable in the marginal statistics of the other two measurements.

The expression above suggests that a simpler form is possible, obtained by preparing the system to actually be in a operational eigenstate of $M_1$, with the value $Q_1=+1$.  In this case all $D_{1,q_j,q_k}=0$.  As $P(Q_{1}\!=\!-1)=0$ for any measurements on such an initial preparation, it will also be the case that $P_{(M_1,M_2,M_3)}(Q_{1}\!=\!-1,Q_{2}\!=\!-1,Q_{3}\!=\!-1)=0$ and that $D_{2,-1,q_k}=0$.  This produces
\begin{equation}
\langle Q^+\rangle_{LG} = 2 D_{2,+1,+1}+4 P_{(M_1,M_2,M_3)}(Q_{1}\!=\!+1,Q_{2}\!=\!+1,Q_{3}\!=\!+1)-1
\end{equation}
This simplified inequality applies to an experimental arrangement where the system is prepared in some state, there is a choice of whether measurement $M_2$ takes place, then measurement $M_3$ is always performed.  If the introduction of $M_2$ does not affect the statistics of the $M_3$ measurement, then $D_{2,+1,+1}=0$ and the Leggett-Garg inequality cannot be violated.

\end{appendix}

\section*{Acknowledgements}
We thank George Knee, Erik Gauger, Andrew Briggs, Simon Benjamin, Rob Spekkens, and Guido Bacciagaluppi for helpful discussion. This work was supported by grants from the John Templeton Foundation and the Templeton World Charity Foundation.

%\bibliographystyle{apalike}
%\bibliography{lgbib}

\end{document}